\documentclass[11pt,a4paper]{article}
\pdfoutput=1

\usepackage{jheppub}
\usepackage{latexsym}
\usepackage{revsymb}
\usepackage{multirow}
\usepackage{color}
\usepackage[usenames,dvipsnames,svgnames,table]{xcolor}

\usepackage{graphicx}
\usepackage{epsfig}  
\usepackage{epsf}    
\usepackage{dcolumn}
\usepackage{bm}
\usepackage{dcolumn}
\usepackage{textcomp}
\usepackage{float}
\usepackage{hypcap}
\usepackage[]{hyperref}
\usepackage{adjustbox}
\usepackage{multirow}

\usepackage{amsmath}
\usepackage{subfigure}
\usepackage{alphalph}
\usepackage{morefloats}
\usepackage{textcomp}
\usepackage[utf8x]{inputenc}
\usepackage{float}
\restylefloat{table}

\newcommand{\be}{\begin{equation}}
\newcommand{\ee}{\end{equation}}
\newcommand{\ba}{\begin{eqnarray}}
\newcommand{\ea}{\end{eqnarray}}

\title{Optimal configuration of Protvino to ORCA experiment for hierarchy and non-standard interactions} 

\author{Dinesh Kumar Singha,}
\author{Monojit Ghosh,}
\author{Rudra Majhi,}
\author{Rukmani Mohanta}

\affiliation{School of Physics, University of Hyderabad, Hyderabad - 500046, India}

\emailAdd{dinesh.sin.187@gmail.com}
\emailAdd{monojit$\_$rfp@uohyd.ac.in}
\emailAdd{rudra.majhi95@gmail.com}
\emailAdd{rmsp@uohyd.ac.in}

\abstract{
In this paper, we study the hierarchy sensitivity of Protvino to ORCA (P2O) experiment in three flavour scenario as well as its sensitivity to non-standard interactions (NSI) in neutrino propagation. Because of the largest possible baseline length of 2595 km, P2O is expected to have strong sensitivity towards neutrino mass hierarchy and NSI parameters. In our study, we show that even though the number of appearance channel events for the minimal configuration of P2O are higher compared to DUNE, still the hierarchy sensitivity of P2O is less than DUNE because of large background events. Our results show that for a background reduction factor of 0.46 and appearance channel background systematic normalization error of 4\%, the hierarchy sensitivity of P2O becomes equivalent of DUNE for $\delta_{\rm CP} = 195^\circ$. We call this configuration of P2O as optimized P2O.
 Regarding the study of NSI, we find that, for $\epsilon_{e\mu}$ ($\epsilon_{e\tau}$) sensitivity of DUNE is similar (better) as compared to optimized P2O when both $\epsilon_{e\mu}$ and $\epsilon_{e\tau}$ are included in the analysis. Our results show that in presence of NSI, the change of hierarchy sensitivity with respect to standard three flavor scenario, is higher in P2O as compared to DUNE. Further, hierarchy sensitivity in presence of NSI is lower (higher) than sensitivity in the standard three flavour scenario for $\delta_{\rm CP} = 270^\circ (90^\circ)$. It is important to note that hierarchy sensitivity of optimized P2O  does not get significantly better than DUNE for the current favourable values of $\delta_{\rm CP}$ which is $180^\circ < \delta_{\rm CP} < 360^\circ$ as obtained by the global analysis in both standard three flavour and in presence of NSI.
}

\arxivnumber{}
\begin{document}
\maketitle

\section{Introduction}
\label{intro}
The main motivation of the future generation neutrino oscillation experiments is to precisely determine the unknowns in the standard three flavour scenario and to probe the existence of new physics  beyond it. In the standard three flavour scenario, neutrino oscillation is described by three mixing angles: $\theta_{12}$, $\theta_{13}$ and $\theta_{23}$, two mass squared differences $\Delta m^2_{21}$ and $\Delta m^2_{31}$, and one Dirac type CP phase $\delta_{\rm CP}$. Among these parameters the unknowns at this moment are: (i) mass hierarchy of the neutrinos which can be either normal i.e., $\Delta m^2_{31} > 0$ or inverted i.e., $\Delta m^2_{31} < 0$, (ii) the octant of the atmospheric mixing angle which can be either upper i.e., $\theta_{23} > 45^\circ$ or lower i.e., $\theta_{23} < 45^\circ$ and (iii) the CP phase $\delta_{\rm CP}$ \cite{Gonzalez-Garcia:2021dve}. Apart from the standard three flavour scenario, neutrino oscillation experiments are also sensitive to non-standard neutrino interactions (NSI) in neutrino propagation \cite{Ohlsson:2012kf,Miranda:2015dra,Farzan:2017xzy}. In the standard neutral-current (NC) interactions with matter, the initial and the final flavour of the neutrinos remain same during neutrino propagation. However, if there exists non-standard neutral-current interactions, then the initial and the final states of the neutrinos can be different. Existence of such kind of interactions modify the neutrino oscillation probabilities. Therefore, future neutrino oscillation experiments can put bound on these NSI parameters. Further, NSI parameters can also alter the sensitivities of the future experiments for the determination of the unknowns of the standard three flavour scenario. Note that charged-current (CC) non-standard interactions can also affect the production and detection of the neutrinos. However, the parameters of CC NSI are strongly constrained as compared to the NC NSI parameters \cite{Biggio:2009nt} and such NSI's are relevant for the low energy experiments. Therefore, in our study we will  consider the effect of NC NSI only.

The Protvino to ORCA (P2O) \cite{Akindinov:2019flp} and the Deep Underground Neutrino Experiment (DUNE) \cite{DUNE:2020ypp} are the future accelerator based long-baseline neutrino oscillation experiments which aim to probe the earth matter effect to determine one of the major unknowns of the standard three flavour scenario i.e., the neutrino mass hierarchy. The determination of neutrino mass hierarchy depends on the matter effect and therefore longer baseline ensures the measurement of neutrino mass hierarchy with higher confidence level. P2O experiment will have the highest possible baseline of 2595 km whereas DUNE will have a baseline of 1300 km. Interestingly, the baseline length of P2O experiment is very close the bi-magic baseline \cite{Raut:2009jj,Dighe:2010js}, which allows the determination of mass hierarchy by resolving the hierarchy - $\delta_{\rm CP}$ degeneracy \cite{Barger:2001yr,Prakash:2012az,Ghosh:2015ena}. Further, the NSI parameters  also depend on the matter effect and therefore these experiments are capable of putting very strong bounds on the NSI parameters. Measurement of hierarchy sensitivity in standard three flavour scenario \cite{Barger:2013rha,Barger:2014dfa,Deepthi:2014iya,Ghosh:2014rna,Fukasawa:2016yue,DeRomeri:2016qwo,Chakraborty:2017ccm,Ballett:2016daj,Rout:2020emr,Rout:2020cxi,DUNE:2020ypp}  and study of NSI \cite{Masud:2015xva,deGouvea:2015ndi,Coloma:2015kiu,Liao:2016hsa,Masud:2016bvp,C:2016nrg,Coloma:2016gei,Masud:2016nuj,Blennow:2016etl,Agarwalla:2016fkh,Blennow:2016jkn,Fukasawa:2016lew,Deepthi:2016erc,Liao:2016orc,Ghosh:2017ged,Masud:2017bcf,Ghosh:2017lim,Deepthi:2017gxg,Meloni:2018xnk,Flores:2018kwk,Verma:2018gwi,Masud:2018pig,Liao:2019qbb,DUNE:2020fgq,Bakhti:2020fde,Chatterjee:2021wac} in the context of DUNE has been extensively performed in the past. The estimation of hierarchy sensitivity in the context of P2O for three flavour scenario has been carried out in Refs. \cite{Choubey:2018rnl,Akindinov:2019flp,Kaur:2021rau} and a brief study of NSI for P2O can be found in \cite{Feng:2019mno}.
In Ref. \cite{Akindinov:2019flp}, the sensitivity of P2O has been studied for the three configurations: (i) minimal configuration of ORCA detector with 90 KW beam, (ii) updated accelerator configuration of 450 KW beam with ORCA detector and (iii) updated accelerator configuration of 450 KW beam with the updated Super-ORCA detector. In this paper, we consider the minimal configuration of the ORCA detector with 90 KW beam. 

First, we will show that though the baseline of P2O and the corresponding number of expected $\nu_e$ events are higher than DUNE, the hierarchy sensitivity of P2O is less than DUNE in the standard three flavour scenario. We identify that this is because of the very large background of the P2O experiment. Then we find the optimal configuration for P2O in terms of background and systematic errors for which the hierarchy sensitivity of P2O becomes comparable to the sensitivity of DUNE. We use this optimized configuration of P2O, to calculate the bounds on the NSI parameters and compare the sensitivity with original configuration of P2O and DUNE assuming there is no NSI in nature. Next, assuming that NSI exists in nature, we estimate the hierarchy sensitivity of optimized P2O, P2O and DUNE in presence of NSI.

The paper is organized as follows. In the next section, we will briefly discuss the configuration of P2O and DUNE that we use in our calculation. In Section \ref{prob} we revisit the bi-magic baseline condition and show how the bi-magic condition of P2O can help in determining neutrino mass hierarchy without the hierarchy - $\delta_{\rm CP}$ degeneracy. In Section \ref{hier}, we discuss the hierarchy sensitivity of P2O and compare with DUNE in standard three flavour scenario. In Section \ref{nsi}, we present the sensitivity of P2O and DUNE in presence of NSI. Finally in Section \ref{sum}, we will summarize our findings and conclude. In the appendix we will give the event spectrum and hierarchy sensitivity of P2O which we reproduced following Ref. \cite{Akindinov:2019flp}.

\section{Experimental setup and simulation details}
\label{spec}

We simulate both the experiments P2O and DUNE using the GLoBES software \cite{Huber:2004ka,Huber:2007ji}. For implementing NSI probability engine, we use the new physics plugin provided by GLoBES \cite{Kopp:2006wp}. 

For simulating P2O, we have used the configuration as given in Ref. \cite{Akindinov:2019flp}. The U-70 synchrotron located at Protvino, Russia will produce a 90 KW beam corresponding to $0.8 \times 10^{20}$ protons on target per year. These neutrinos will be detected at the ORCA detector located at the Mediterranean Sea 40 km offshore Toulon, France. The distance between the neutrino source and the detector will be around 2595 km. The detector will be a total of 8 Mt of sea water. We have used the fluxes as given in the Fig. 5 of Ref. \cite{Akindinov:2019flp} and matched the energy spectrum as given in Fig. 7 using pre-smearing energy dependent efficiencies (shown in appendix). These efficiencies take care the energy dependent effective mass of the ORCA detector as given in the Figs. 90 of Ref. \cite{KM3Net:2016zxf}. We have taken the energy resolution and particle identification factor from Fig. 68 and 99 of Ref. \cite{KM3Net:2016zxf} respectively. We have adopted the systematic error from Table. 1 of Ref. \cite{Akindinov:2019flp}. We list the value of systematic error in the second column of Table \ref{table_sys}. For estimating the overall systematic error for background normalization error and shape error, we have added the relevant systematic errors of Table. 1 of Ref. \cite{Akindinov:2019flp} in quadrature. We have considered a total run-time of 6 years divided into 3 years in neutrino mode and 3 years in antineutrino mode. 

For DUNE, we have used the official GLoBES files of the DUNE technical design report~\cite{DUNE:2021cuw}. A 40~kt liquid argon time-projection chamber detector 
is placed 1300~km from the source having a power of 1.2~MW 
delivering $1.1 \times 10^{21}$ protons on target per year with a running time of 7~years. We have divided the run-time into 3.5 years in neutrino mode and 3.5 years in antineutrino mode. The neutrino source will be located at Fermilab, USA and the detector will be located at South Dakota, USA. We have listed the systematic errors for DUNE in the 4th coloumn of  Table \ref{table_sys}. Note that DUNE GLoBES file does not contain any shape errors.  

\begin{table} 
\centering
\begin{tabular}{|c|c|c|c|} \hline
Systematics            & P2O         & P2O Optimized               & DUNE  \\ \hline
Sg-norm $\nu_{e}$  & 5$\%$   & 5$\%$ & 2$\%$      \\ 
Sg-norm $\nu_{\mu}$ & 5$\%$  & 5$\%$                     & 5$\%$ \\ 
Bg-norm $\nu_e$ & 12$\%$     & 4$\%$             & 5$\%$ to 20$\%$\\ 
Bg-norm $\nu_\mu$ & 12$\%$    & 12$\%$         & 5$\%$ to 20$\%$\\ 
Bg-shape       & 11$\%$     &  11$\%$            & NA\\ 
Sg-shape       & 11$\%$     &  11$\%$            & NA\\ 
\hline
\end{tabular}
\caption{The values of systematic errors that we considered in our analysis. ``norm" stands for normalization error, ``Sg" stands for signal and ``Bg" stands for background.}
\label{table_sys}
\end{table}  

For the estimation of the sensitivity we use the Poisson log-likelihood and assume that it is $\chi^2$-distributed
\begin{equation}
 \chi^2_{{\rm stat}} = 2 \sum_{i=1}^n \bigg[ N^{{\rm test}}_i - N^{{\rm true}}_i - N^{{\rm true}}_i \log\bigg(\frac{N^{{\rm test}}_i}{N^{{\rm true}}_i}\bigg) \bigg]\,,
\end{equation}
where $N^{{\rm test}}$ is the number of events in the test spectrum, $N^{{\rm true}}$ is the number of events in the true spectrum and $i$ is the number of energy bins. The systematic is incorporated by the method of pull \cite{Fogli:2002pt,Huber:2002mx}. The best-fit values of the oscillation parameters and the $3 \sigma$ values of $\theta_{23}$ are adopted from NuFIT \cite{Esteban:2020cvm} and we list them in Table \ref{table_param}. We marginalize the phase $\delta_{\rm CP}$ in the full range. We have also marginalized $\sin^2 \theta_{23}$ as mentioned in the above table.  We have kept the rest of the parameters fixed in  both true and test spectra of the $\chi^2$. We show all our results for the normal hierarchy of the neutrino masses. 

\begin{table} 
\centering
\begin{tabular}{|c|c|c|} \hline
Parameters            & True values               & Test value Range  \\ \hline
$\sin^2 \theta_{12}$  & 0.304 & NA      \\ 
$\sin^2 \theta_{13}$ & 0.02221                     & NA \\ 
$\sin^2 \theta_{23} $ & 0.57                  & $0.4\rightarrow 0.62$\\ 
$\delta_{\rm CP} $       & $ 195^\circ$                  & $0^\circ \rightarrow 360^\circ $\\ 
$\Delta m^2_{12}$    & $7.42 \times 10^{-5}~{\rm eV}^2 $ & 
NA \\ 

$\Delta m^2_{31}$    &~ $ 2.514 \times 10 ^{-3}~{\rm  eV}^2$ (NH)~~&~ NA~ \\
 \hline
\end{tabular}
\caption{The values of oscillation parameters that we considered in our analysis.}
\label{table_param}
\end{table}  


\section{Bi-magic property of the P2O baseline}
\label{prob}

The appearance channel ($\nu_\mu \rightarrow \nu_e$) probability for neutrinos in presence of matter can be expressed in terms of the small parameters $\alpha = \Delta m^2_{21}/\Delta m^2_{31}$ and $s_{13}$ as \cite{Akhmedov:2004ny}

 \begin{align}
 \label{ap_prob}
  P_{\mu e} & = 4 s^{2}_{13}s^{2}_{23}\frac{\sin^2{[(1 -\hat A)\Delta]}}{(1-\hat A)^2} + \alpha^{2} \cos^{2}\theta_{23} \sin^{2}2 \theta_{12} \frac{\sin^{2}\hat A\Delta}{\hat A^2} 
  \\ \nonumber & +\alpha s_{13} \sin 2\theta_{12}  \sin 2\theta_{23}\cos(\Delta+\delta_{\rm CP}) \frac{\sin{\hat A \Delta}}{\hat A} \frac{\sin{[(1-\hat A)\Delta]}}{(1-\hat A)}, 
 \end{align}
 where, $\Delta = \Delta m^2_{31}L/4E$, $s_{ij} (c_{ij}) \equiv \sin{\theta_{ij}}(\cos{\theta_{ij}})$ and $\hat{A} = 2\sqrt{2} G_F n_e E / \Delta m^2_{31}$, $G_F$ is the Fermi constant, $n_e$ is the electron number density, $E$ is the energy of neutrino and $L$ is the baseline length. For neutrinos, $\hat{A}$ is positive for normal hierarchy and negative for inverted hierarchy, while for antineutrinos it is the opposite. Furthermore, for antineutrinos $\delta_{\rm CP} \rightarrow -\delta_{\rm CP}$.  At the oscillation maximum, $\Delta$ corresponds to $90^\circ$ in vacuum. In presence of matter term, the oscillation maximum will be slightly shifted from $\Delta = 90^\circ$. However, for simplicity we will assume $\Delta = 90^\circ$ to draw our conclusions from the anlytic expressions. From the above probability expression, we understand that for neutrinos, $\delta_{\rm CP} = 270^\circ$ corresponds to the maximum point in the probability and $\delta_{\rm CP} = 90^\circ$ corresponds to the minimum point in the probability. This is opposite for antineutrinos. Further, for neutrinos the probabilities for normal hierarchy is higher as compared to the probabilities for the inverted hierarchy because of the matter term $\hat A$. This is also opposite for antineutrinos. The separation of the appearance channel probabilities in normal hierarchy and inverted hierarchy provides the hierarchy sensitivity of an experiment. 

\begin{figure}[t]
\begin{center}
\includegraphics[width=0.49\textwidth]{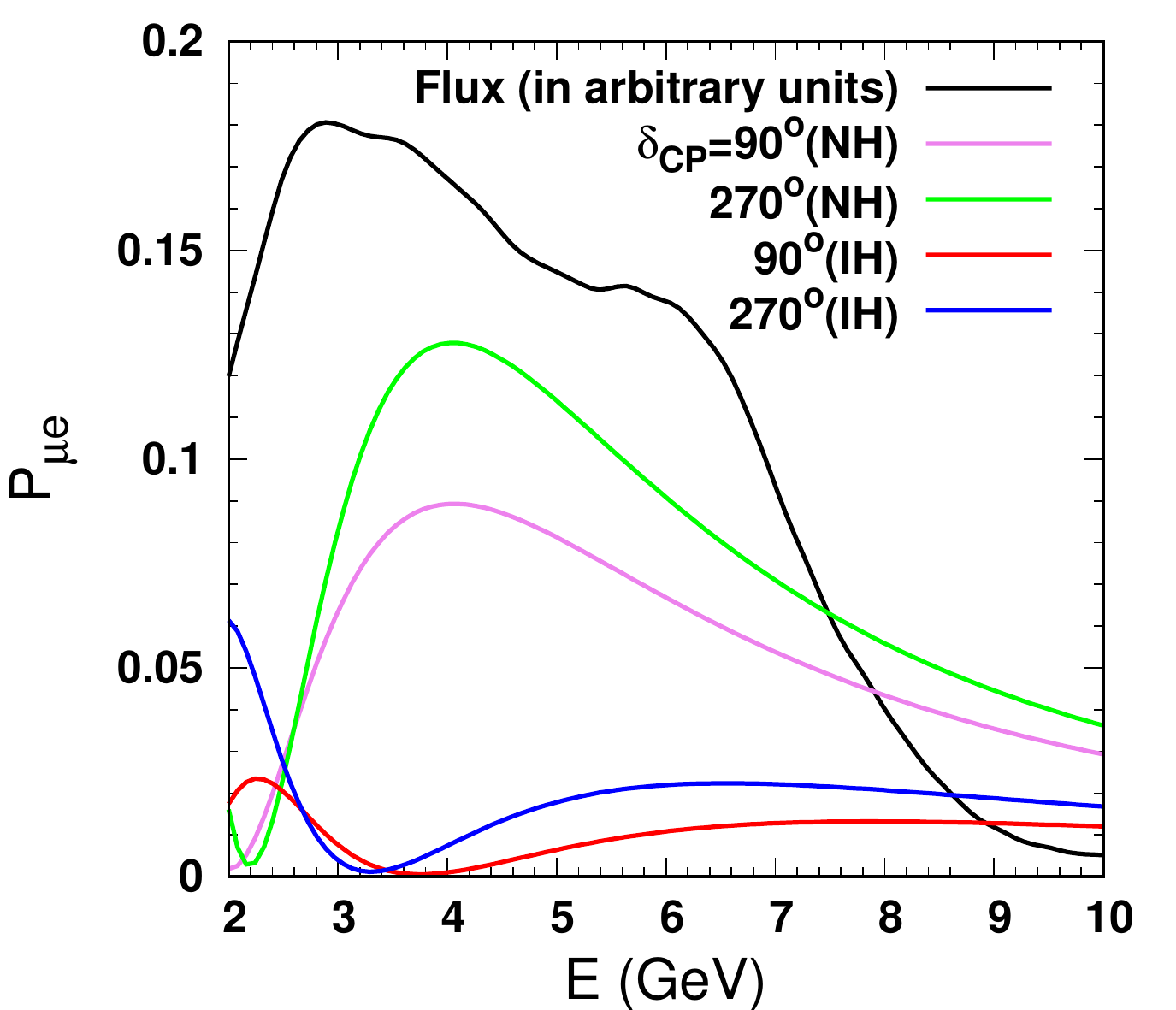} 
\includegraphics[width=0.49\textwidth]{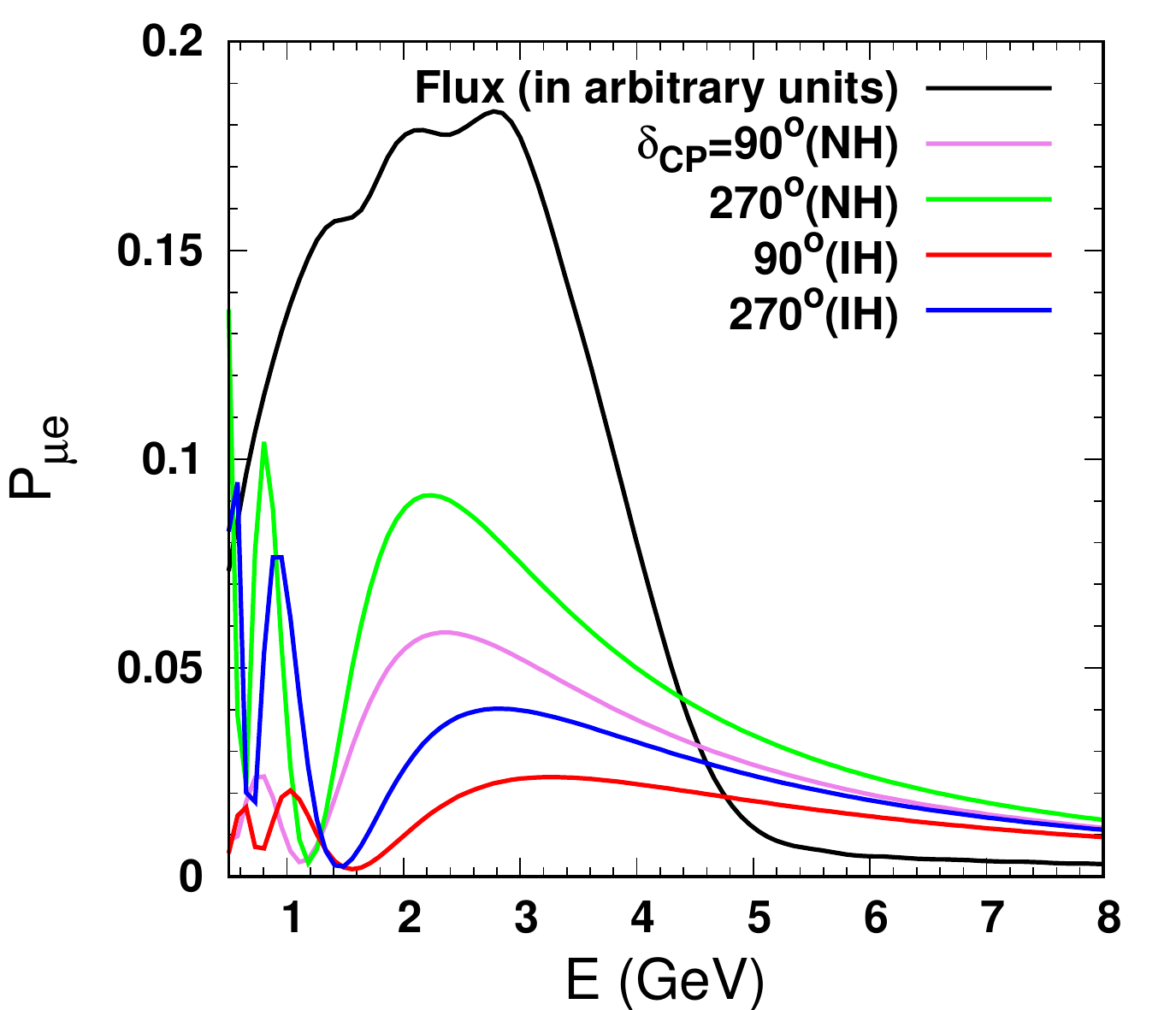} \\
\includegraphics[width=0.49\textwidth]{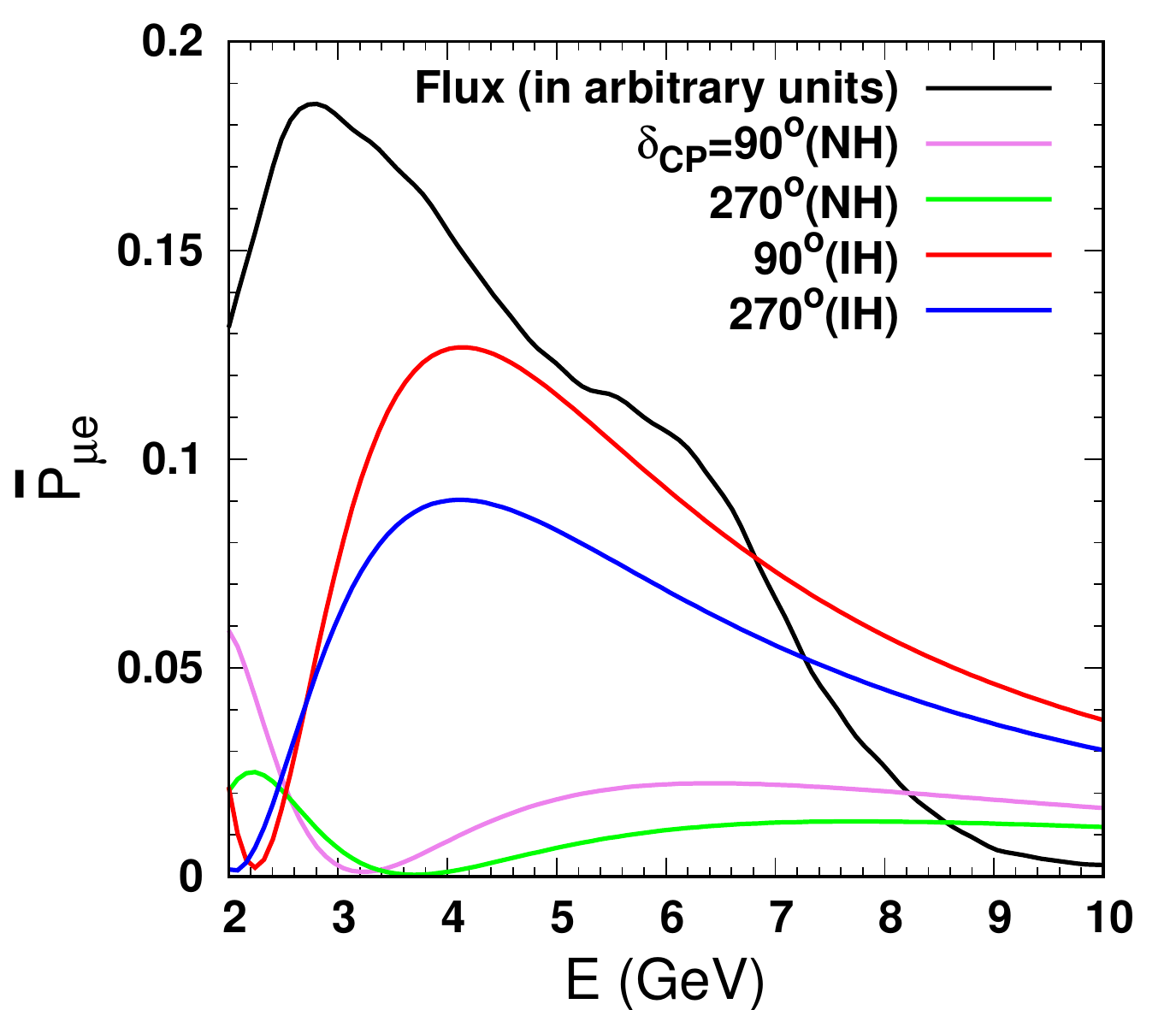} 
\includegraphics[width=0.49\textwidth]{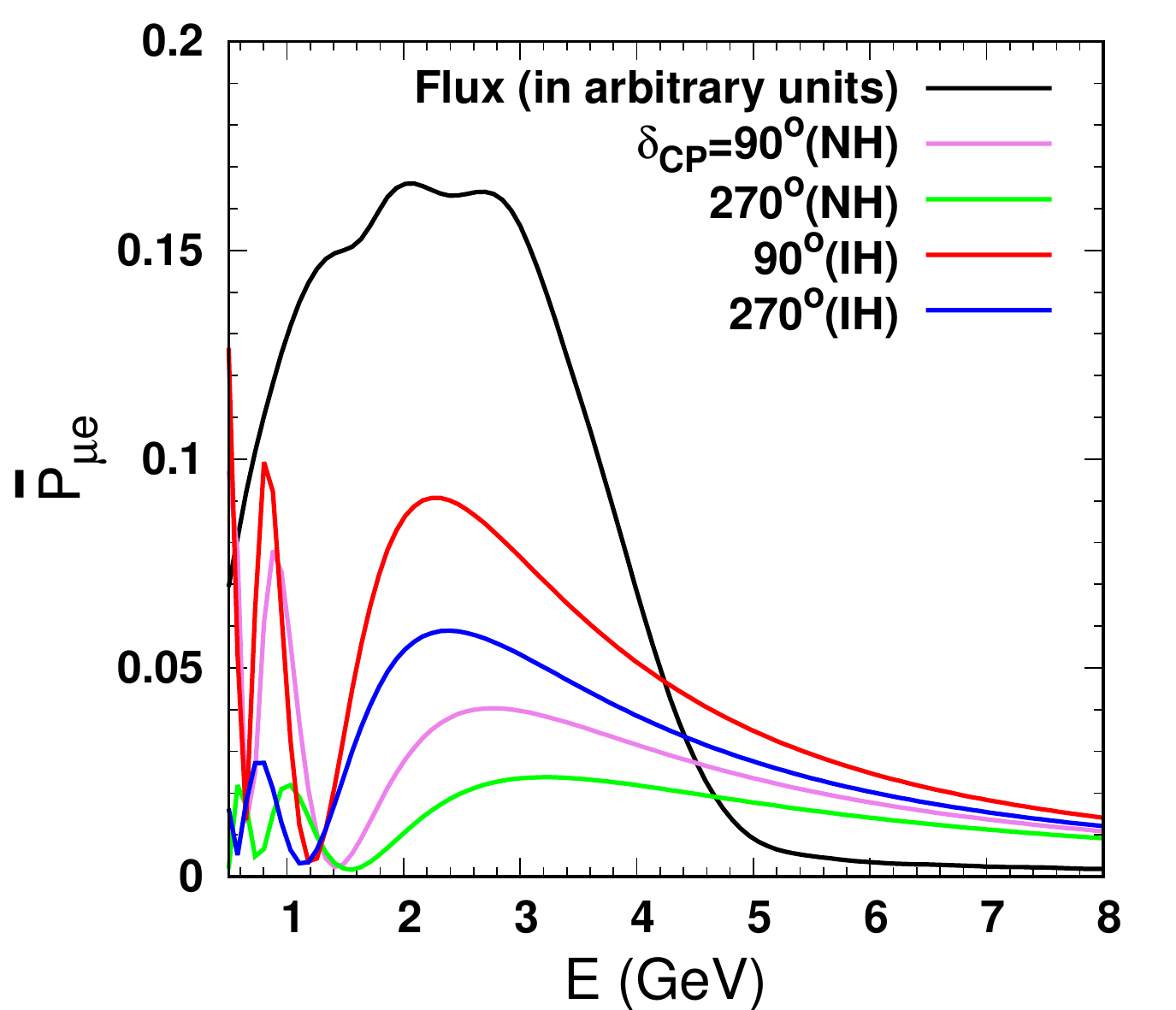}
\end{center}
\caption{Appearance channel probability and flux as a function of energy. The left column is for P2O baseline and the right column is for DUNE baseline. In each column the top panel is for neutrinos and the bottom panel is for antineutrinos.}
\label{fig_prob}
\end{figure}

In Fig. \ref{fig_prob}, we have plotted the appearance channel probability as a function of neutrino energy. The left column is for P2O baseline and the right column is for DUNE baseline. In each column the top panel is for neutrinos and the bottom panel is for antineutrinos. In generating these panels we have used the value of the oscillation parameters as given in Table \ref{table_param}. In each panel, green/purple/blue/red corresponds to the ($\delta_{\rm CP}$, hierarchy) combinations of ($270^\circ$, NH)/($90^\circ$, NH)/($270^\circ$, IH)/($90^\circ$, IH) respectively. The black curve shows the corresponding fluxes of the experiments P2O and DUNE. The region covered by the black curve is the energies to which these experiments are sensitive to. In these panels we clearly realize the behaviour of the curves which we discussed in the previous paragraph. From these panels we also understand that the separation between the NH and IH curves are higher for the P2O baseline as compared to the DUNE baseline due to higher matter effect of the former.

It was shown that for $L \sim 2540$ km, there is no $\delta_{\rm CP}$ dependence in IH and at the same time there is a probability maximum in NH at 3.3 GeV \cite{Raut:2009jj,Dighe:2010js}. 
This baseline is known as the bi-magic baseline. As the baseline of P2O is around 2595 km, this bi-magic property can also be visible in Fig. \ref{fig_prob}. From the top left panel we notice that, around 3.5 GeV, the blue and the red curves intersect to a single point. This implies the fact that the appearance channel probabilities for different values of $\delta_{\rm CP}$ becomes same for inverted hierarchy. In other words, at this value of energy, the appearance channel probability becomes independent of  $\delta_{\rm CP}$ in the inverted hierarchy. However, for NH, the variation of the probability with respect to $\delta_{\rm CP}$ becomes maximum at $E = 3.5$ GeV. Therefore at this point the separation between the probabilities in the normal hierarchy and the inverted hierarchy becomes largest. We do not observe such features in the oscillation probabilities of DUNE.  

Further,  it was also shown that for $L \sim 2540$ km, there is no $\delta_{\rm CP}$ dependence for NH and a probability maximum in IH at 1.9 GeV. As the flux in that energy is very small, the P2O experiment will be unable to probe that region.

\section{Hierarchy sensitivity in standard three flavour case}
\label{hier}

\begin{figure}[t]
\begin{center}
\includegraphics[width=0.49\textwidth]{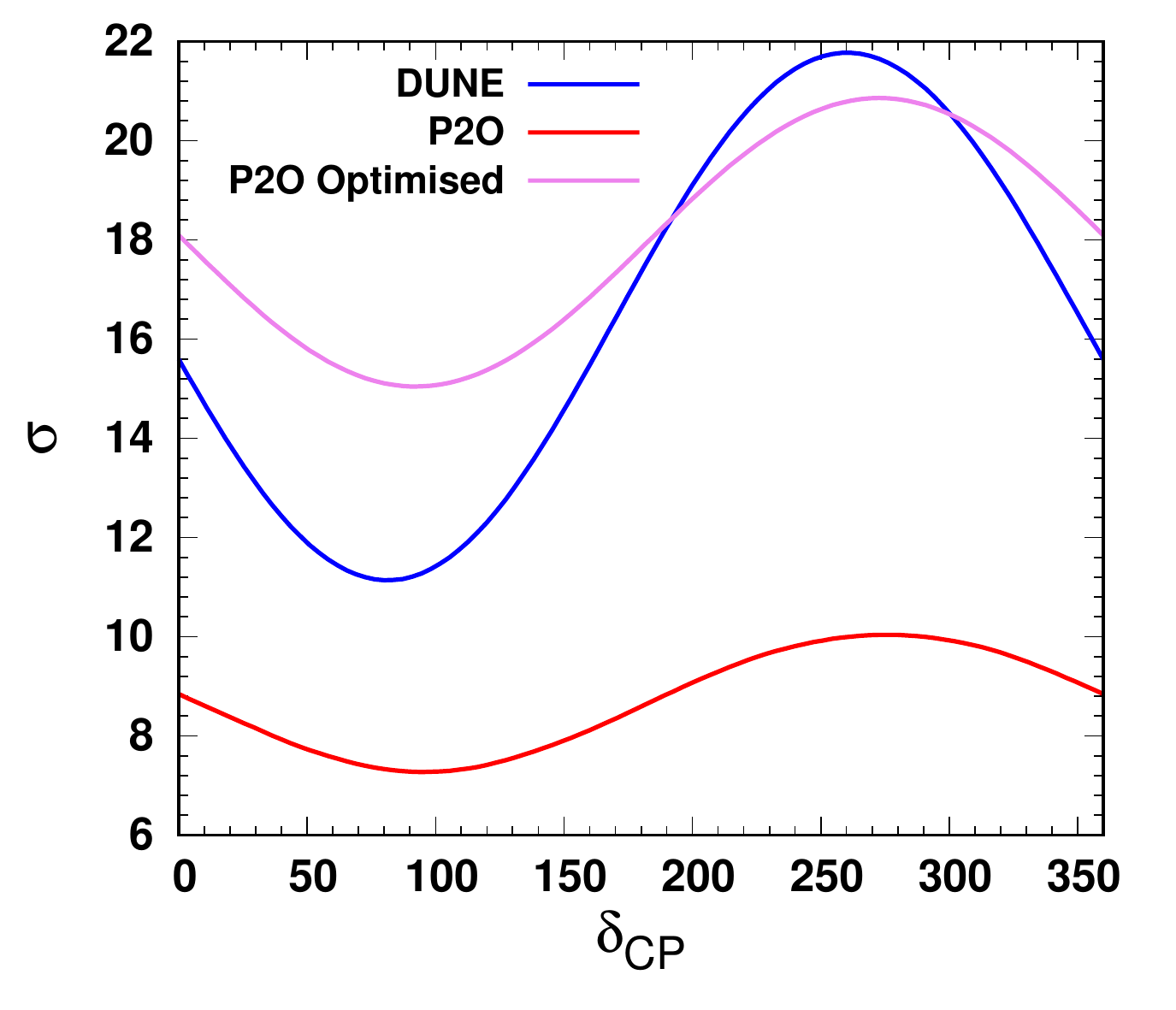}
\includegraphics[width=0.49\textwidth]{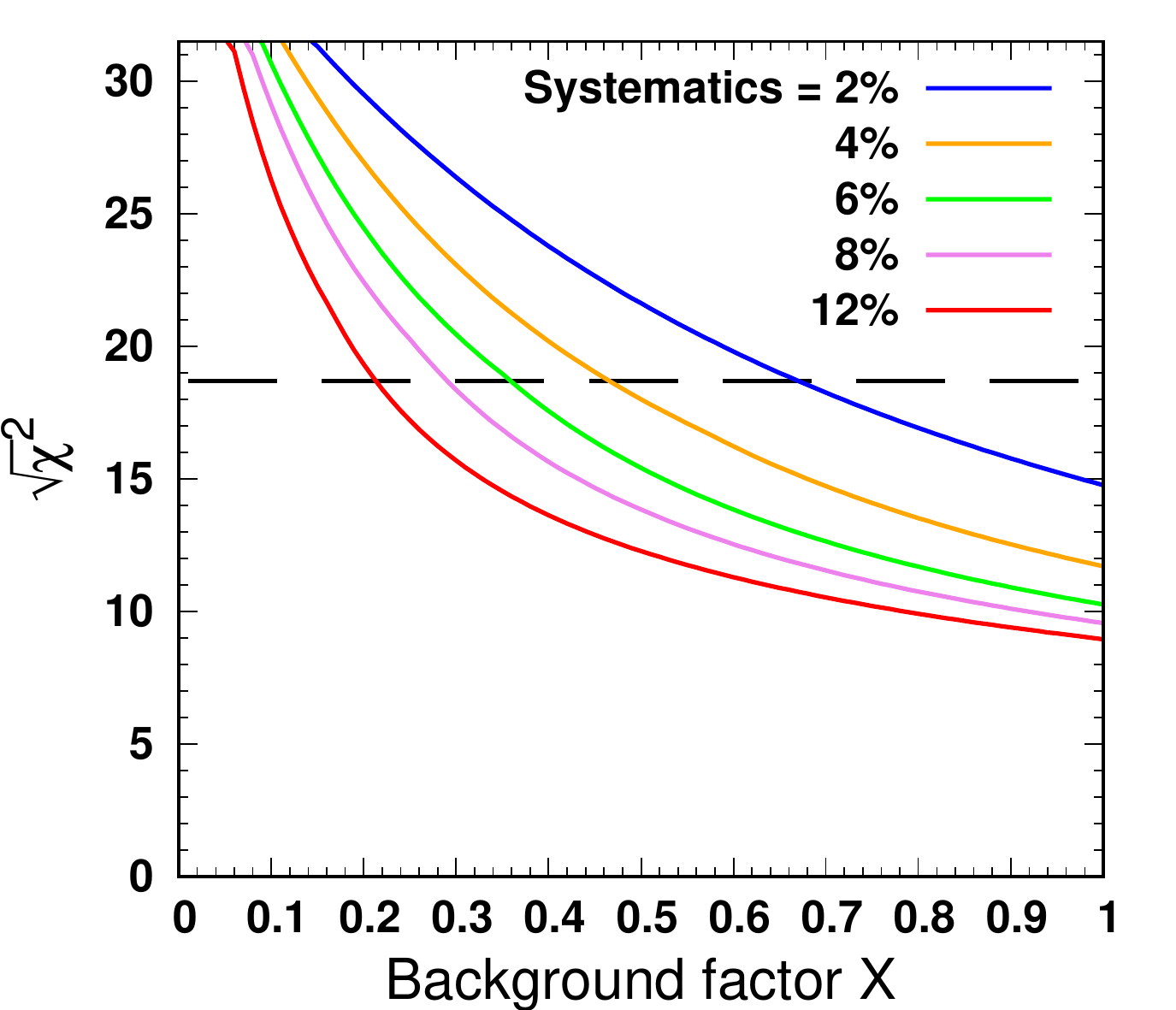} \end{center}
\caption{Hierarchy sensitivity as a function of true $\delta_{\rm CP}$ (left panel) and as a function of background reduction factor $X$ (right panel). These panels are generated for true normal hierarchy.}
\label{hier_stan}
\end{figure}

In the left panel of Fig. \ref{hier_stan}, we have plotted the hierarchy sensitivity as a function of true $\delta_{\rm CP}$. The blue curve corresponds to DUNE and the red curve corresponds to P2O. We have checked that our hierarchy sensitivity of P2O matches with the sensitivity as given in Ref. \cite{Akindinov:2019flp} (shown in appendix). From the panel we see that the hierarchy sensitivity is maximum for true value of $\delta_{\rm CP}$ around $270^\circ$ and minimum for $\delta_{\rm CP}$ value around $90^\circ$ for both P2O and DUNE. This can be understood by looking at Fig. \ref{fig_prob}. Hierarchy sensitivity is proportional to the separation between the NH curves and the blue curves in IH. From all the panels of Fig. \ref{fig_prob}, we realize that the curve corresponding to $\delta_{\rm CP}$ around $270^\circ$ in NH (green curve) is more separated from the blue IH curve as compared to separation between the blue IH curve and the curve corresponding to $\delta_{\rm CP}$ around $90^\circ$ (purple curve) in NH. Further we also note that the variation of hierarchy sensitivity with respect to true $\delta_{\rm CP}$ is higher in DUNE and lower in P2O. This can also be understood from Fig. \ref{fig_prob}. From the antineutrino probabilities in Fig. \ref{fig_prob} (bottom row), we see that for P2O around $3.5$ GeV, where the probabilities corresponding to all the values of $\delta_{\rm CP}$ are same, the separation of $\delta_{\rm CP} = 270^\circ$ and $90^\circ$ in NH (which is basically same point) from the blue curve are same whereas for DUNE, as mentioned earlier, the green curve is more separated from the blue curve as compared to the separation between the blue curve and the purple curve. Therefore, when sensitivities from neutrino channel and antineutrino channels are combined, the variation of the hierarchy sensitivity with respect to $\delta_{\rm CP}$ becomes higher in DUNE as compared to P2O.

The most important thing that we note from the left panel of Fig. \ref{hier_stan} is that though the baseline of P2O is higher than DUNE and P2O baseline has the bi-magic property, the sensitivity of P2O is much lower as compared to the sensitivity of DUNE. To understand that we have calculated the total number $\nu_e$ events for both DUNE and P2O for both the hierarchies. We have presented these numbers in Table \ref{table_events_sg}. These events correspond to the 3 years running of P2O and 3.5 year running of DUNE. 
\begin{table}
\centering
\begin{tabular}{ |p{3cm}|p{3cm}|p{3cm}|  }
\hline
\multicolumn{3}{|c|}{$\nu_{e}$ events} \\
\hline
Experiments& NH &IH \\
\hline
P2O& 2404 & 528 \\
DUNE & 1380 & 702 \\
\hline
\end{tabular}
\caption{Total number of $\nu_e$ events for P2O and DUNE. These events corresponds to 3 years running of P2O and 3.5 year running of DUNE. }
\label{table_events_sg}
\end{table}
\begin{table}
\centering
\begin{tabular}{ |p{3cm}|p{3cm}|p{3cm}|p{3cm}|  }
\hline
\multicolumn{4}{|c|}{Background Events (NH) (for appearance channel)}  \\
\hline
Experiments& $\nu_{\mu}$ &NC & $\nu_{\tau}$ \\
\hline
P2O& 2166 & 1235 & 873 \\
DUNE & 24 & 87 & 45 \\
\hline
\end{tabular}
\caption{Total number of background events for the $\nu_e$ channel for P2O and DUNE. These events corresponds to 3 years running of P2O and 3.5 year running of DUNE.}
    \label{table_events_bg}
\end{table}
As the hierarchy sensitivity of the long-baseline experiments come from the appearance channel, the difference between the $\nu_e$ events in NH and IH is proportional to the hierarchy sensitivity of an experiment. From this table we understand that the difference between $\nu_e$ events in NH and IH are higher in P2O as compared to DUNE. Therefore, we expect to have higher hierarchy sensitivity in DUNE as compared to P2O but in reality that is not the case. To understand this, in Table \ref{table_events_bg}, we have calculated the total number of background events for the $\nu_e$ selection for P2O and DUNE in NH. From this table, we understand that the backgrounds of P2O corresponding to selection of $\nu_\mu$ as $\nu_e$, NC as $\nu_e$ and $\nu_\tau$ as $\nu_e$  are very high as compared to DUNE. The ORCA detector for P2O mainly designed to detect atmospheric neutrinos and ultra high energy neutrinos from the extra-galactic sources. The electron events in the ORCA detectors produce a shower. From the right panel of Fig. 99 of Ref. \cite{KM3Net:2016zxf}, we see that at 5 GeV, the efficiency of $\nu_e$ events classified as shower is 90\% which we consider as signal. But there are also 78\% of the $\nu_\tau$ events, 80\% of the NC events and 50\% of the $\nu_\mu$ events are classified at shower which act as a background for the $\nu_e$ selection. Whereas for the DUNE detector, which is designed for detection of the neutrinos from the accelerator, this background efficiencies are small. Therefore, we understand that because of the limited background rejection capability of P2O, the hierarchy sensitivity of this experiment is smaller than DUNE.

There are two ways to improve the sensitivity of P2O without increasing the beam power. Either to reduce the background efficiencies or to improve the background systematic error of the appearance channel. This can be achieved in the the more densely-instrumented version of the ORCA detector known as Super-ORCA.
With the Super-ORCA detector, one can achieve lower energy threshold for neutrino detection, better neutrino flavour identification capability and better energy resolution compared to ORCA. For Super-ORCA, the energy threshold for neutrino detection will be reduced to 0.5 GeV, the selection efficiency of muon-like and electron-like events will be 95\% and the neutrino energy resolution is $\sim 20\%$ at $E > 1$ GeV \cite{Akindinov:2019flp}.
 In the right panel of Fig. \ref{hier_stan}, we have plotted the hierarchy sensitivity as a function of a background reduction factor $X$ for different values of background systematic normalization error for the appearance channel. We have generated these curves for the current best-fit value of $\delta_{\rm CP}$ of $195^\circ$. In this panel the black dashed horizontal line corresponds to the sensitivity of DUNE. The factor $X$ is basically a constant by which we multiply all the background efficiencies that we mention in the previous paragraph. The intersection of different curves with the black dashed curve shows the combination of $X$ and background systematic error for which the sensitivity of P2O becomes equal to DUNE. From this panel we see that for $X = 0.46$ and systematic error of 4\%, the sensitivity of P2O becomes equivalent to DUNE. Therefore from now, we call this configuration of P2O as optimized P2O. In third column of Table \ref{table_sys}, we list the corresponding systematic errors for the optimized P2O.

The hierarchy sensitivity of optimized P2O is represented by the purple curve in the left panel of Fig. \ref{hier_stan}. Here we notice that the sensitivity of optimized P2O is better than DUNE for $\delta_{\rm CP} = 90^\circ$ and slightly worse than DUNE for $\delta_{\rm CP} = 270^\circ$. 
Therefore we see that the hierarchy sensitivity of the optimized P2O is equivalent (better) as compared to DUNE for $180^\circ (0^\circ) < \delta_{\rm CP} < 360^\circ (180^\circ)$ which is currently favoured (disfavoured) by the global analysis.
 In the next section we will study the sensitivity of P2O, optimized P2O and DUNE in presence of NSI. 


\section{Sensitivity in presence of NSI}
\label{nsi}

The NC NSI in neutrino propagation can arise from the following four-fermion interaction:
\begin{eqnarray}
{\cal L}_{\mbox{\rm\scriptsize eff}}^{\mbox{\tiny{\rm NSI}}} 
=-2\sqrt{2}\, \epsilon_{\alpha\beta}^{ff'P} G_F
\left(\overline{\nu}_{\alpha L} \gamma_\mu \nu_{\beta L}\right)\,
\left(\overline{f}_P \gamma^\mu f_P'\right),
\label{NSIop}
\end{eqnarray}
where $f_P$ and $f_P'$ correspond to fermions with chirality $P$ and
$\epsilon_{\alpha\beta}^{ff'P}$ is a dimensionless constant. In the presence of NSI, the matter term $\hat A$ is modified by the following factor:
\begin{eqnarray}
\left(
\begin{array}{ccc}
1+ \epsilon_{ee} & \epsilon_{e\mu} & \epsilon_{e\tau}\\
\epsilon_{\mu e} & \epsilon_{\mu\mu} & \epsilon_{\mu\tau}\\
\epsilon_{\tau e} & \epsilon_{\tau\mu} & \epsilon_{\tau\tau}
\end{array}
\right),
\label{matter-np}
\end{eqnarray}
where $\epsilon_{\alpha\beta}$ is defined by
\begin{equation}
\epsilon_{\alpha\beta}\equiv\sum_{f=e,u,d}\frac{N_f}{N_e}\epsilon_{\alpha\beta}^{f}\,.
\end{equation}
$N_f~(f=e, u, d)$ is the number densities of fermions $f$. Here we defined the NSI parameters as $\epsilon_{\alpha\beta}^{fP}\equiv\epsilon_{\alpha\beta}^{ffP}$ and
$\epsilon_{\alpha\beta}^{f}\equiv\epsilon_{\alpha\beta}^{fL}+\epsilon_{\alpha\beta}^{fR}$. The present $90\%$ bounds of the NSI parameters are given by\,\cite{Davidson:2003ha,Biggio:2009nt} 
\begin{eqnarray}
\hspace{-25pt}
&{\ }&
\left(
\begin{array}{lll}
|\epsilon_{ee}| < 4 \times 10^0 & |\epsilon_{e\mu}| < 3\times 10^{-1}
& |\epsilon_{e\tau}| < 3 \times 10^0\\
&  |\epsilon_{\mu\mu}| < 7\times 10^{-2}
& |\epsilon_{\mu\tau}| < 3\times 10^{-1}\\
& & |\epsilon_{\tau\tau}| < 2\times 10^1
\end{array}
\right).
\label{epsilon-m}
\end{eqnarray}

In our analysis we will consider the off-diagonal complex NSI parameters $\epsilon_{e \mu} = |\epsilon_{e \mu}|e^{i\phi_{e\mu}}$ and $\epsilon_{e\tau} = |\epsilon_{e \tau}|e^{i\phi_{e\tau}}$. Our choice of NSI parameters are motivated by the following. Recently in Refs. \cite{Chatterjee:2020kkm,Denton:2020uda}, it was shown that the discrepancy between the $\delta_{CP}$ measurement in the experiments T2K \cite{T2K:2021xwb} and NO$\nu$A \cite{NOvA:2021nfi} can be resolved by introducing the NSI parameters $\epsilon_{e \mu}$ and $\epsilon_{e\tau}$. Taking one-parameter at a time, the discrepancy was resolved for the values of $|\epsilon_{e \mu}| = 0.15~(0.19)$, $\phi_{e\mu} = 250^\circ~(270^\circ)$ and  $|\epsilon_{e \tau} | = 0.27~(0.28)$, $\phi_{e\tau} = 290^\circ~(288^\circ)$ in Ref. \cite{Chatterjee:2020kkm} (\cite{Denton:2020uda}). Therefore, in this work we will study the sensitivity of P2O and DUNE in presence of the NSI parameters $\epsilon_{e \mu}$ and $\epsilon_{e \tau}$. First we will study the capability of these experiments to exclude these values of NSI parameters if there exist no NSI in Nature. In addition, we will also study how the hierarchy sensitivity of these experiments change in presence of NSI. In our analysis we will take the best-values of Ref. \cite{Chatterjee:2020kkm} as a reference point.

\subsection{Constraining the NSI parameters $\epsilon_{e\mu}$ and $\epsilon_{e\tau}$ }

\begin{figure}[t]
\begin{center}
\includegraphics[width=0.49\textwidth]{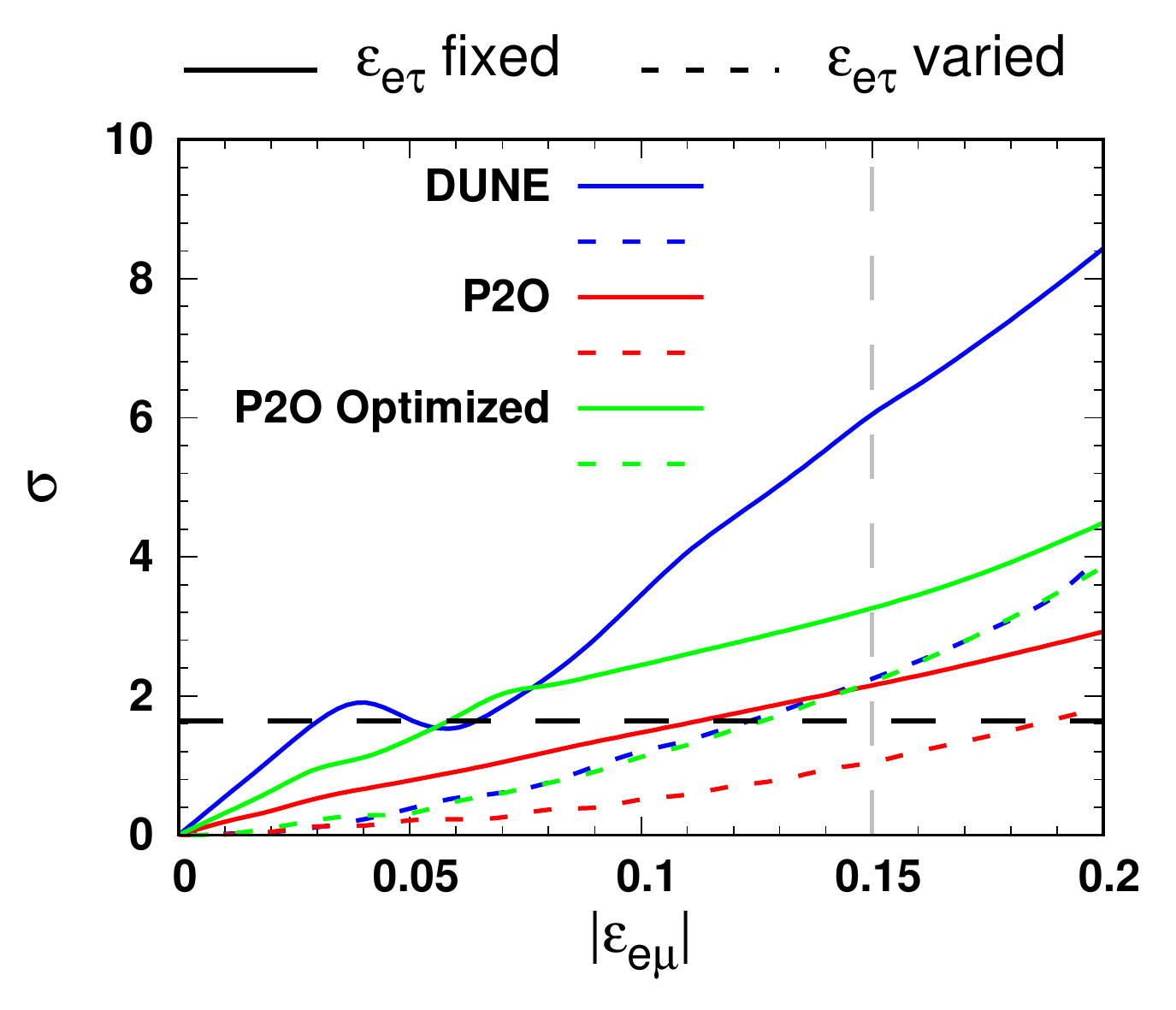}
\includegraphics[width=0.49\textwidth]{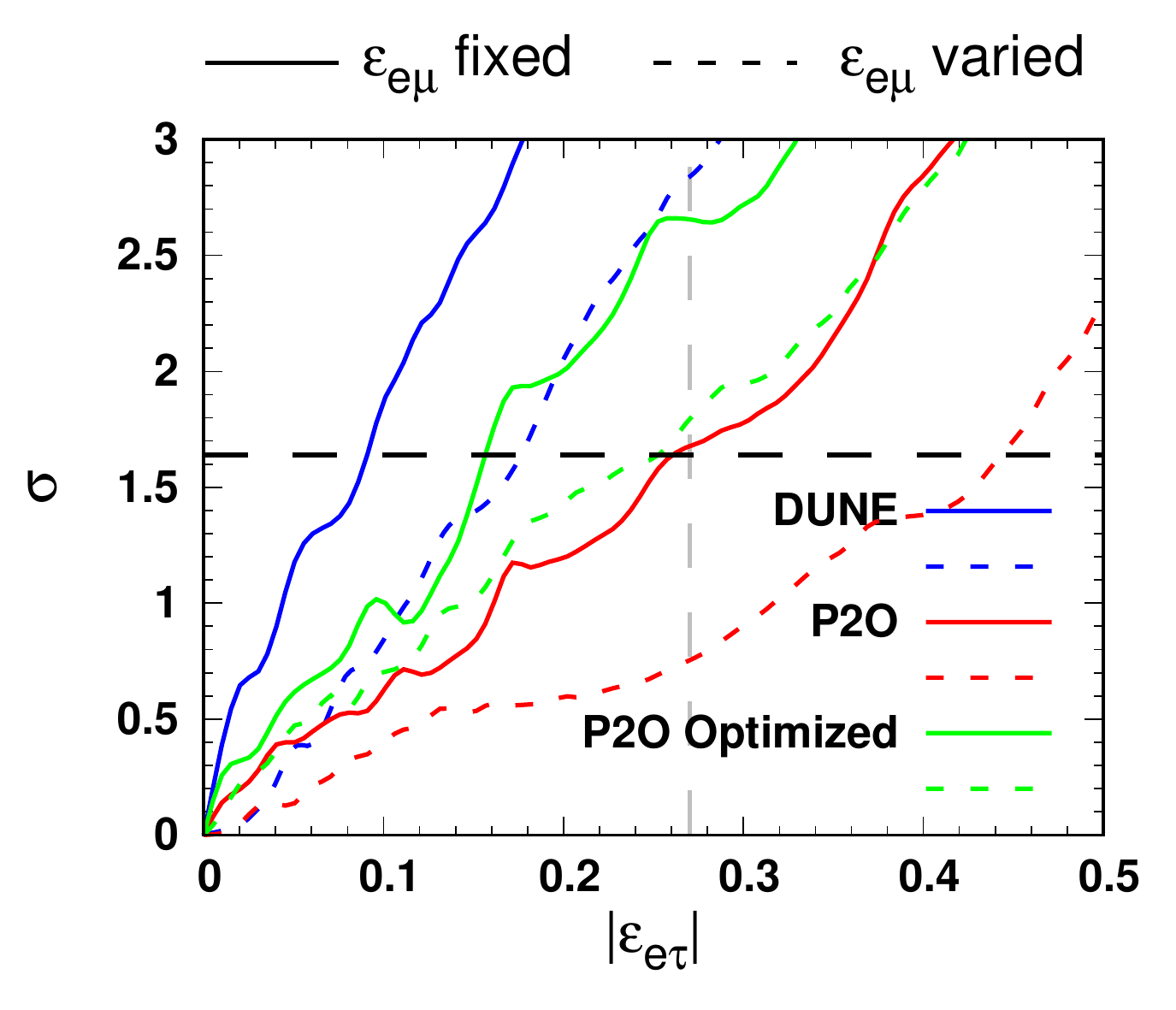} \\
\includegraphics[width=0.49\textwidth]{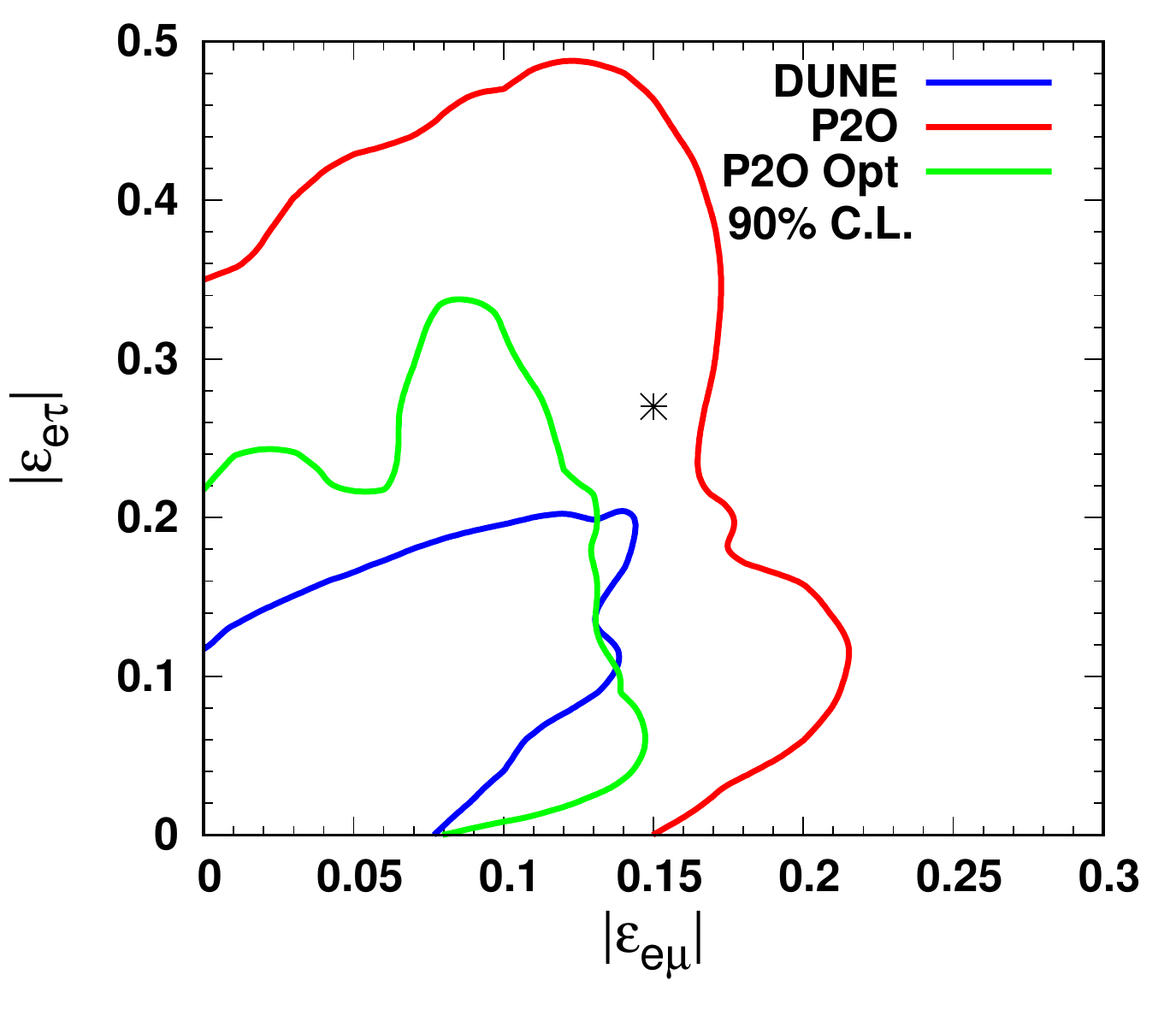}
\end{center}
\caption{Capability of P2O and DUNE to constrain the NSI parameters. In the top left (top right) panel we present the one-dimensional sensitivity curves for $|\epsilon_{e \mu}|$ ($|\epsilon_{e\tau}|$). In the bottom panel we present 90\% contours in the $|\epsilon_{e \mu}|$ - $|\epsilon_{e\tau}|$ plane. These panels are generated for true normal hierarchy.}
\label{fig_nsi}
\end{figure}

In Fig. \ref{fig_nsi}, we have presented the capability of P2O and DUNE to constrain the NSI parameters $\epsilon_{e\mu}$ and $\epsilon_{e\tau}$. The top left panel corresponds to the one-dimensional sensitivity curves for $|\epsilon_{e \mu}|$ and the top right panel corresponds to the one-dimensional sensitivity curves for $|\epsilon_{e \tau}|$. Bottom panel corresponds to 90\% contours in the $|\epsilon_{e \mu}|$ - $|\epsilon_{e\tau}|$ plane. In each panel the blue curve corresponds to DUNE, the red curve corresponds to P2O and the green curve corresponds to optimized P2O. In the top panels, the solid curves corresponds to the case when only one-parameter is taken at a time whereas the dashed lines shows the sensitivity when both the parameters are considered in the analysis. The black dashed horizontal line shows the value of $\sigma$ corresponding to 90\% C.L. The grey dashed vertical line reflects that value of $|\epsilon_{e\mu}| = 0.15$ (left panel) and $|\epsilon_{e\tau}| = 0.27$ (right panel) which is the best-fit value of the NSI parameters as obtained in Ref. \cite{Chatterjee:2020kkm}. In the bottom panel this is shown by a ``star".  In generating these plots we have calculated the true spectrum of the $\chi^2$ without NSI and minimized the relevant NSI parameters in the test which are not shown in the figure. From these panels we note that the capability of P2O and DUNE to constrain $\epsilon_{e\mu}$ is better than $\epsilon_{e\tau}$. This is because of the fact that in the accelerator experiments, the bounds on NSI are obtained from a muon beam. From the top panels we also note that the sensitivity becomes stronger when we consider one NSI parameter at a time as compared to the case when both the NSI parameters are included in the analysis. For $\epsilon_{e\mu}$ (top left panel), we note that sensitivity of DUNE is still better than optimized P2O if we do not include $\epsilon_{e\tau}$ in the analysis whereas they becomes similar if we include $\epsilon_{e\tau}$ in the analysis. However, for $\epsilon_{e\tau}$ (top right panel) the sensitivity of DUNE is always better than optimized P2O irrespective of inclusion of $\epsilon_{e\mu}$ in the analysis. These conclusions can also be observed from the bottom panel where both $|\epsilon_{e\mu}|$ and $|\epsilon_{e\tau}|$ are included in the analysis. The upper bound on $|\epsilon_{e\mu}|$ at 90\% C.L. obtained from DUNE is similar as that of optimized P2O whereas the upper bound on $|\epsilon_{e\tau}|$ at 90\% C.L. obtained from DUNE is lower than optimized P2O. We present the 90\% limit on these NSI parameters obtained from our analysis in Table. \ref{table_nsi}.
\begin{table}
\centering
\begin{tabular}{ |p{3cm}|p{3cm}|p{3cm}|  }
\hline
\multicolumn{3}{|c|}{90\% bound on the NSI parameters} \\
\hline
Experiments& $|\epsilon_{e\mu}|$ &$|\epsilon_{e\tau}|$  \\
\hline
P2O& 0.112 (0.188) & 0.26 (0.444) \\
Optimized P2O & 0.058 (0.126) & 0.157 (0.28) \\
DUNE & 0.065 (0.123) & 0.09 (0.176) \\
\hline
\end{tabular}
\caption{90\% bound on the NSI parameters. The numbers in the parenthesis corresponds to the case when both $\epsilon_{e\mu}$ and $\epsilon_{e\tau}$ are included in the analysis.}
\label{table_nsi}
\end{table}
From our analysis we see that the values $|\epsilon_{e\mu}| = 0.15$ and $|\epsilon_{e\tau}| = 0.27$ will be excluded at 90\% C.L. by all the experiments irrespective of the values of $\phi_{e \mu}$ and $\phi_{e \tau}$ except P2O when both $\epsilon_{e\mu}$ and $\epsilon_{e\tau}$ are included in the analysis. 

\begin{figure}
\begin{center}
\includegraphics[width=0.75\textwidth]{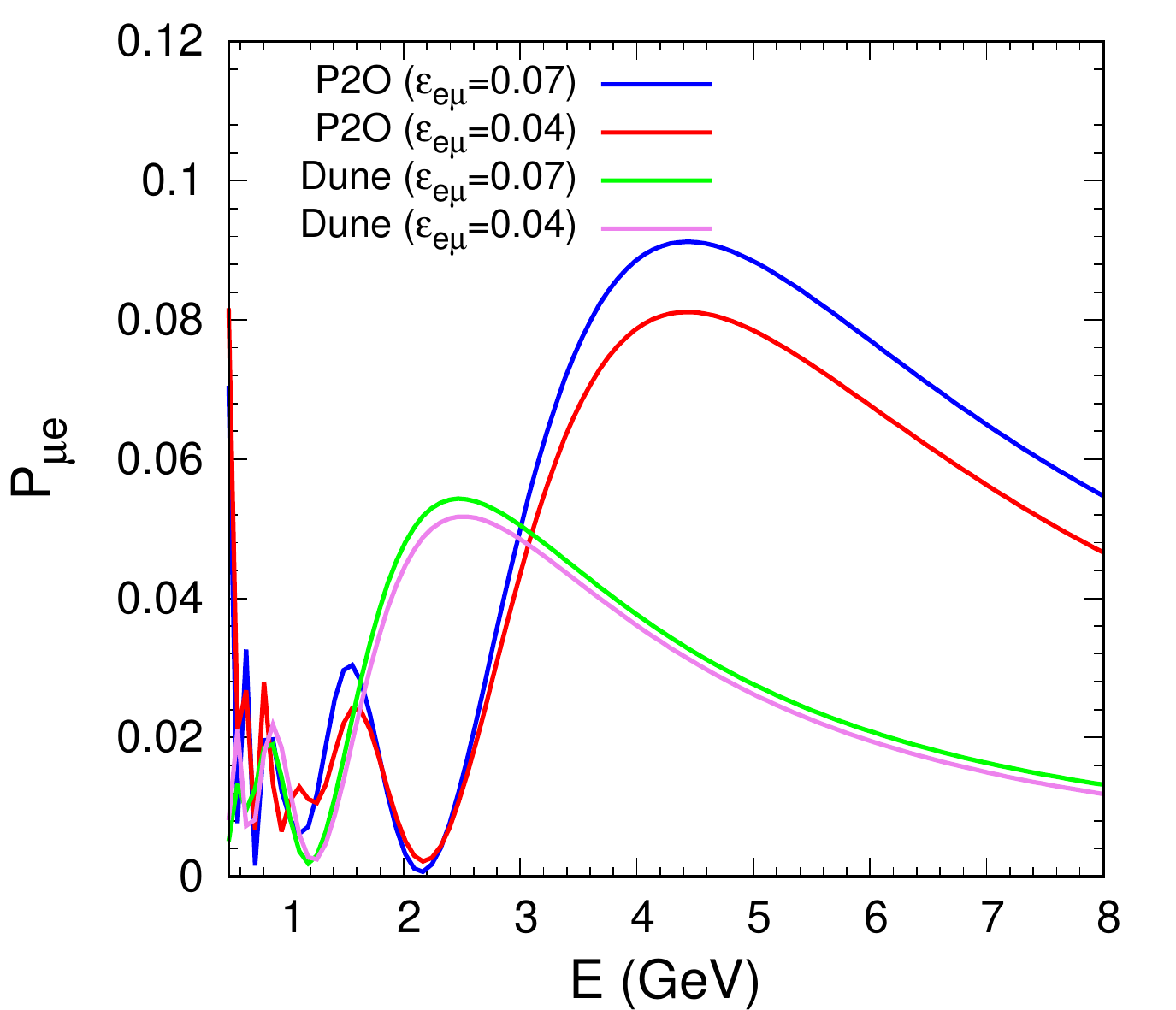}
\end{center}
\caption{Appearance channel probability for neutrinos considering two sets of parameters in P2O and DUNE baselines. The sets of parameters are: $|\epsilon_{e \mu} = 0.04|$, $\sin^2\theta_{2 3} = 0.46$, $\phi_{e \mu} = 210^\circ$, $\delta_{\rm CP} = 60^\circ $ and  $|\epsilon_{e \mu}| = 0.07$, $\sin^2\theta_{2 3} = 0.45$, $\phi_{e \mu} = 195^\circ$, $\delta_{\rm CP} = 75^\circ $. This panel is generated for true normal hierarchy.}
\label{prob_deg}
\end{figure}

\begin{figure}[hbt]
\begin{center}
\includegraphics[width=0.49\textwidth]{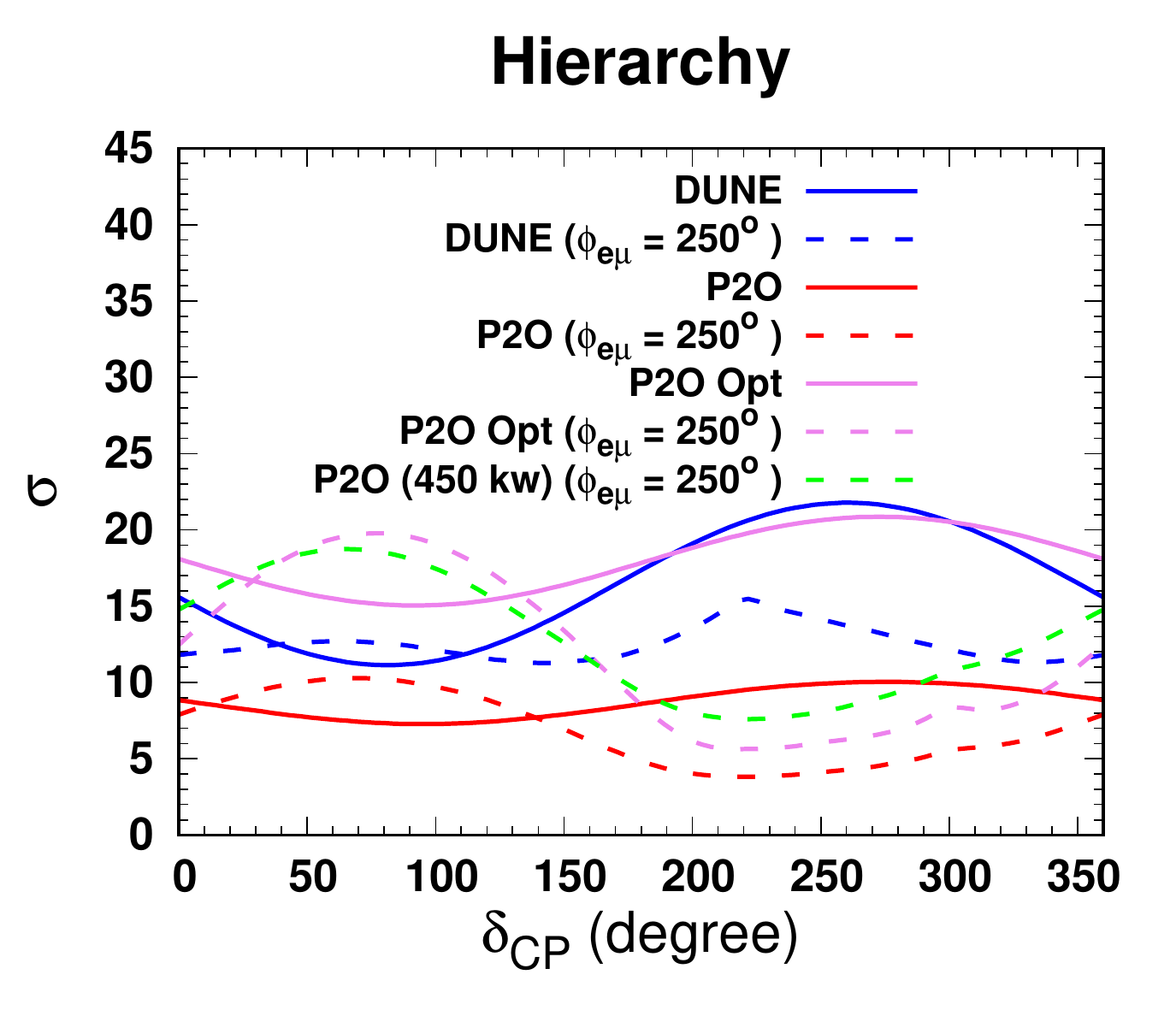}
\includegraphics[width=0.49\textwidth]{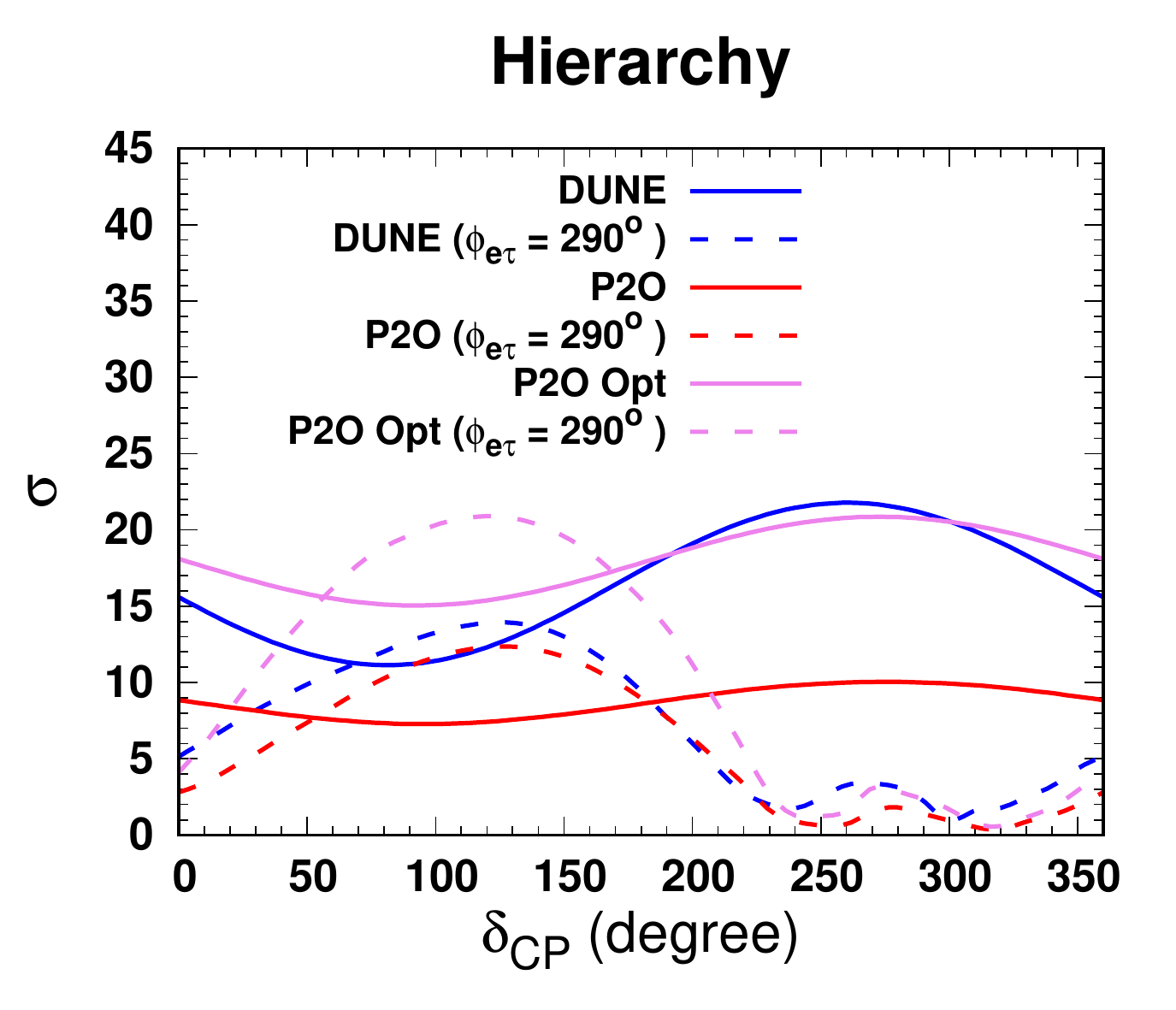}
\includegraphics[width=0.49\textwidth]{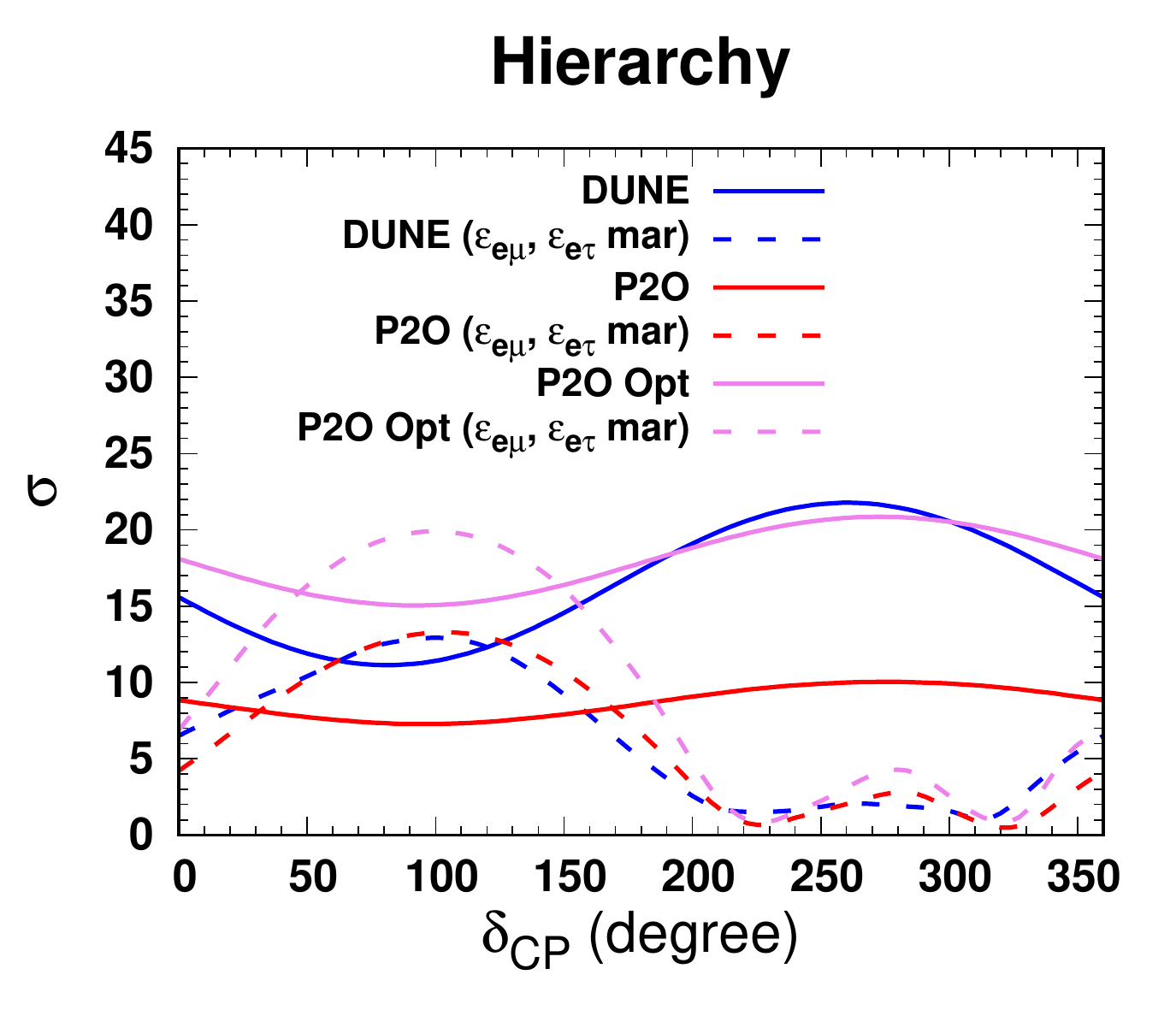}
\end{center}
\caption{Hierarchy sensitivity as a function of $\delta_{\rm CP}$ (true) in presence of NSI. The top left (right) panel is for only $\epsilon_{e \mu}$ ($\epsilon_{e \tau}$). The bottom panel is when both $\epsilon_{e \mu}$ and $\epsilon_{e \tau}$ are included in the analysis. These panels are generated for true normal hierarchy.}
\label{fig_hier_nsi}
\end{figure}

It is interesting to note that in the top left panel, the solid blue curve corresponding to DUNE, does not increase continuously as $|\epsilon_{e\mu}|$ increases. This does not happen for P2O or the case of DUNE when both $\epsilon_{e\mu}$ and $\epsilon_{e\tau}$ are included in the analysis. At $|\epsilon_{e\mu}| = 0.04$, the sensitivity starts to decreases as $|\epsilon_{e\mu}|$ increases and then after $|\epsilon_{e\mu}| = 0.07$, the sensitivity increases again. This happens because of the degeneracy between $|\epsilon_{e\mu}|$, $\theta_{23}$, $\delta_{\rm CP}$ and $\phi_{e \mu}$ in DUNE when only $\epsilon_{e\mu}$ is taken in the analysis. To understand this, first we find out at the test values of oscillation parameters which are minimized in the $\chi^2$ i.e., $\theta_{23}$ (test), $\delta_{\rm CP}$ (test) and $\phi_{e \mu}$ (test) at $|\epsilon_{e\mu}| = 0.04$ and $|\epsilon_{e\mu}| = 0.07$ for which the $\chi^2$ becomes minimum. These values are $\sin^2\theta_{23} = 0.46$, $\phi_{e \mu} = 210^\circ$, $\delta_{\rm CP} = 60^\circ$ for $|\epsilon_{e\mu}| = 0.04$ and $\sin^2\theta_{2 3} = 0.45$, $\phi_{e \mu} = 195^\circ$, $\delta_{\rm CP} = 75^\circ$ for $|\epsilon_{e\mu}| = 0.07$. A degeneracy in these two sets of parameters implies, the appearance channel probability for this two sets of points must be very close for DUNE but separated for P2O. This can be seen from Fig. \ref{prob_deg} where we have plotted the appearance channel probability for neutrinos considering these two sets of parameters in P2O and DUNE baselines. From the figure we see that for DUNE, the probability for these two sets of parameters are very close to each other (pink and green curves) where they are separated for P2O (blue and red curves). The larger separation between the blue and the red curve can be attributed to the larger matter effect in P2O as compared to DUNE. This is the reason why the sensitivity of P2O increases gradually as $|\epsilon_{e\mu}|$ increases from $0.04$ to $0.07$ whereas for DUNE it decreases first and then increases. Note that this degeneracy gets resolved in DUNE, when both $\epsilon_{e\mu}$ and $\epsilon_{e\tau}$ are included in the analysis. 

\subsection{Hierarchy sensitivity in presence of $\epsilon_{e \mu}$ and $\epsilon_{e \tau}$}

In Fig. \ref{fig_hier_nsi}, we have presented the hierarchy sensitivity of P2O and DUNE in presence of NSI. The top left panel is the case when only one parameter $\epsilon_{e \mu}$ is considered with true value of $|\epsilon_{e \mu}| = 0.15$ and $\phi_{e\mu} = 250^\circ$. The top right panel is the case when only one parameter $\epsilon_{e \tau}$ is considered with true value of $|\epsilon_{e \tau} | = 0.27$ and $\phi_{e\tau} = 290^\circ$. The bottom panel is when both $\epsilon_{e \mu}$ and $\epsilon_{e \tau}$ are included in the analysis. In generating these panels, we have kept fixed $|\epsilon_{e \mu}|$ and $|\epsilon_{e \tau}|$ in the test spectrum of the $\chi^2$ and minimized over the corresponding phases. In these panels the solid curves are for sensitivity without NSI and the dashed curves are for the sensitivity in presence of NSI. The blue/red/purple curves correspond to DUNE/P2O/optimized P2O. From the panels we note that in presence of NSI, the sensitivity is lower than the sensitivity in the standard three flavour scenario for $\delta_{\rm CP} = 270^\circ$ and higher than the sensitivity in the standard three flavour scenario for $\delta_{\rm CP} = 90^\circ$. Further we see that the change of hierarchy sensitivity due to NSI with respect to standard three flavor oscillation case is higher in P2O as compared to DUNE. This is due to the longer baseline of P2O than DUNE. 

In the introduction we mentioned that there is a possibility to upgrade the accelerator in Protvino to 450 KW. A 450 KW beam running for 3 years in neutrino mode and 3 years in antineutrino mode is equivalent to running of a 90 KW beam for 15 years in neutrino mode and 15 years in antineutrino mode. In the top left panel, the dashed green curve shows the hierarchy sensitivity corresponding to 450 KW beam in presence of NSI. From this panel we understand that the sensitivity corresponding to 450 KW beam is equivalent to our configuration of optimized P2O. 

It is important to note that as in the standard case, in presence of NSI, the hierarchy sensitivity of optimized P2O  does not become better than DUNE for the current favourable values of $\delta_{\rm CP}$ which is $180^\circ < \delta_{\rm CP} < 360^\circ$ as obtained by the global analysis. The sensitivity of optimized P2O becomes only better than DUNE for the unfavourable values of $0^\circ < \delta_{\rm CP} < 180^\circ$. Though we have shown the results for a particular choice of ($|\epsilon_{e \mu}|$, $\phi_{e \mu}$) and ($|\epsilon_{e\tau}|$, $\phi_{e \tau}$), we have checked that this conclusion remains true for all the other values of ($|\epsilon_{e \mu}|$, $\phi_{e \mu}$) and ($|\epsilon_{e\tau}|$, $\phi_{e \tau}$) allowed by the current constraints.

In the top right and bottom panel we see that the hierarchy sensitivity becomes almost zero around $\delta_{\rm CP} = 240^\circ$ in presence of NSI for all the three experiments. This is because of a hierarchy degeneracy which occurs between $\delta_{\rm CP} = 240^\circ$ in NH and ($\delta_{\rm CP}$, $\phi_{e \tau}$) in IH for neutrinos. This degeneracy was absent in the case of standard three flavour scenario and arises because of the existence of NSI \cite{Liao:2016hsa}. 
We will explain this in detail in the next paragraph. 

\begin{figure}[hbt]
\begin{center}
\includegraphics[width=0.49\textwidth]{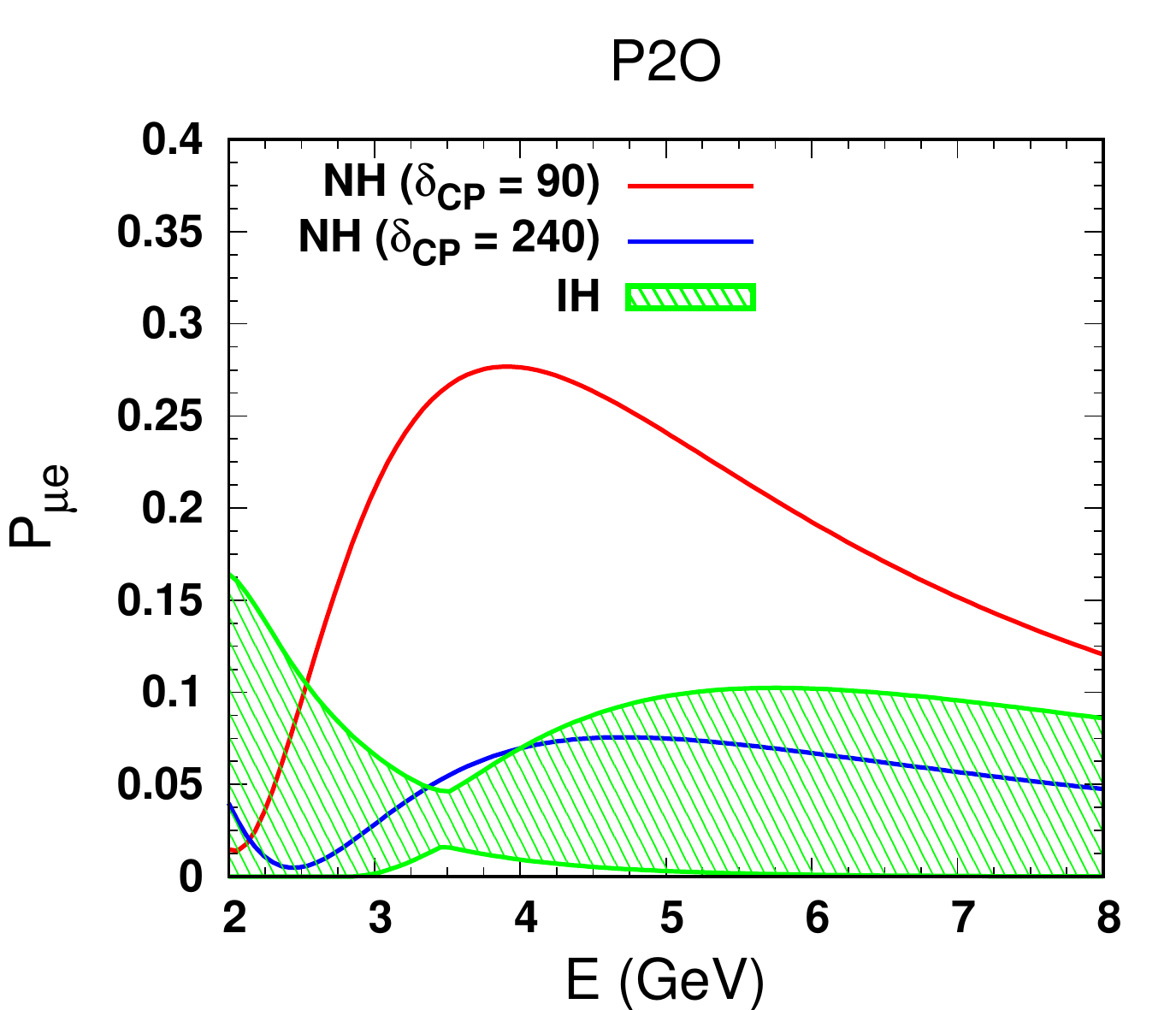}
\includegraphics[width=0.49\textwidth]{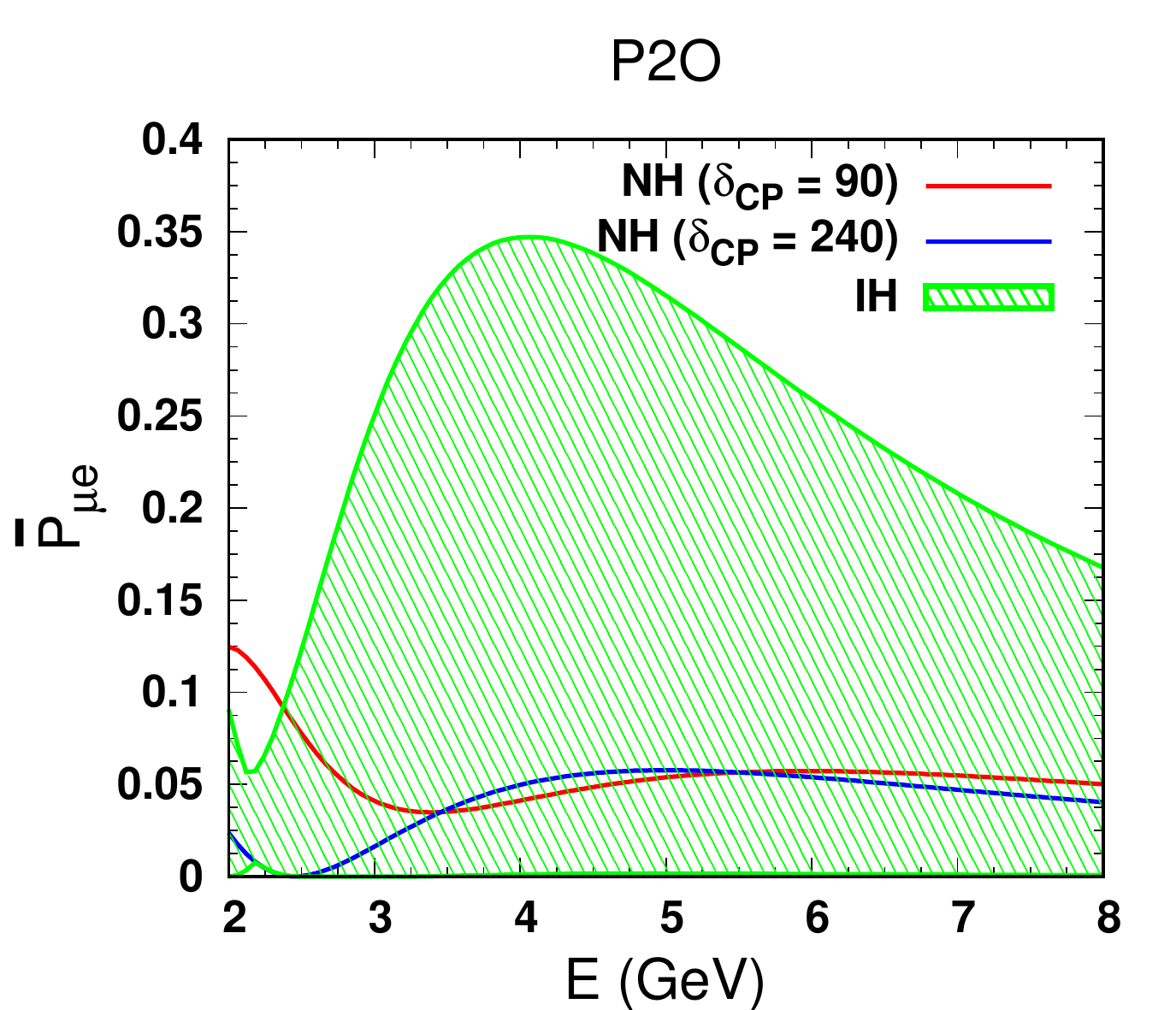}
\end{center}
\caption{Appearance channel probability for neutrinos (left panel) and antineutrinos (right panel) in the presence of NSI parameter $|\epsilon_{e \tau}|=0.27$. The NH curves are generated for $\phi_{e\tau} = 290^\circ$, and $\delta_{\rm CP} = 90^\circ$ and $240^\circ$. In IH these phases are varied.}
\label{fig_prob_nsi}
\end{figure}

The appearance channel formula for neutrinos in presence of $\epsilon_{e \tau}$ takes the following form:
\begin{eqnarray}
P_{\mu e} &=& x^2 f^2 + 2xyfg \cos(\Delta + \delta_{\rm CP}) + y^2 g^2 \\ \nonumber
  &+& 4\hat A \epsilon_{e\tau} s_{23} c_{23} \{xf [f \cos(\phi_{e\tau}+\delta_{\rm CP})
- g \cos(\Delta+\delta_{\rm CP}+\phi_{e\tau})]  \\ \nonumber
&-& y g [g \cos\phi_{e\tau} - f \cos(\Delta-\phi_{e\tau})]\}  + 4 \hat A^2 g^2 c_{23}^2  s_{23}^2\epsilon_{e\tau}^2 \\ \nonumber
 &+&  4 \hat A^2 f^2 s_{23}^2 c_{23}^2\epsilon_{e\tau}^2
- 8 \hat A^2 f g s_{23}^2 c_{23}^2 \epsilon_{e\tau}^2 \cos\Delta\;,  
\end{eqnarray}
where,
\begin{eqnarray}
x &\equiv& 2 s_{13} s_{23}\,,\quad
y \equiv 2 \alpha s_{12} c_{12} c_{23}\,,
\nonumber\\
f &\equiv& \frac{\sin[\Delta(1 - \hat A)]}{(1 - \hat A)}\ ,~~
g \equiv \frac{\sin(\hat A \Delta)}{\hat A}\;.
\label{eq:define}
\end{eqnarray}

For a fixed values of the mixing angles, the value of the probability will be determined by value of the different phases. Therefore, at the probability maximum, we can rewrite the above formula as:

\begin{eqnarray}
\label{eq_nsi}
P_{\mu e} &=& B - 2xyfg \sin\delta_{\rm CP} + 4\hat A \epsilon_{e\tau} s_{23} c_{23} \{xf [f \cos(\phi_{e\tau}+\delta_{\rm CP}) \\ \nonumber
&-& g \sin(\delta_{\rm CP}+\phi_{e\tau})] - y g [g \cos\phi_{e\tau} + f \sin(\phi_{e\tau})]\}\;.
\end{eqnarray}
where $B = x^2 f^2 + y^2 g^2 + 4 \hat A^2 (g^2 + f^2 ) c_{23}^2  s_{23}^2\epsilon_{e\tau}^2 $. Note that in absence of NSI, the maximum point in the probability will correspond to $\delta_{\rm CP} = 270^\circ$ and the minimum point in the probability will correspond to $\delta_{\rm CP} = 90^\circ$ which we have seen in Fig. \ref{fig_prob}. Now in the top right panel of Fig. \ref{fig_hier_nsi} we have seen that hierarchy sensitivity becomes almost zero for $\delta_{\rm CP}$ around $240^\circ$ and it becomes maximum around $\delta_{\rm CP} = 90^\circ$. This plot is generated for true $|\epsilon_{e \tau}|=0.27$ and true $\phi_{e\tau} = 290^\circ$. To understand that in Fig. \ref{fig_prob_nsi} we have plotted the appearance channel probability for neutrinos (left panel) and antineutrinos (right panel) in the presence of NSI parameter $|\epsilon_{e \tau}|=0.27$ for the P2O baseline. The NH curves are generated for $\phi_{e\tau} = 290^\circ$ whereas in the IH, the band is due to the variation of $\delta_{\rm CP}$ and $\phi_{e\tau}$. 
From the panel we note that the curve for $\delta_{\rm CP} = 240^\circ$ overlaps with the IH probabilities in both neutrino and antineutrino probabilities to give rise to a new hierarchy degeneracy because of which the hierarchy sensitivity is completely lost. However, the curve for $\delta_{\rm CP} = 90^\circ$ overlaps with the IH probabilities in antineutrinos but it is free from degeneracy in neutrinos. Therefore a combination of neutrinos and antineutrinos provide a non-zero hierarchy $\chi^2$. Next we will try to explain this degeneracy from the analytical expressions and 
show that this degeneracy is independent of the baseline length. 

For $\phi_{e \tau} = 270^\circ$ (which is close to $\phi_{e\tau} = 290^\circ$), we can write Eq. \ref{eq_nsi} as:
\begin{eqnarray}
P_{\mu e} &=& B - 2xyfg \sin\delta_{\rm CP} + 4\hat A \epsilon_{e\tau} s_{23} c_{23} \{xf [f \sin\delta_{\rm CP} \\ \nonumber
&-& g \cos\delta_{\rm CP}] + y g f\}\;.
\end{eqnarray}
 If we take $\delta_{\rm CP}=90^\circ$ then $P(\nu_\mu \to \nu_e)$ will become,
 \begin{eqnarray}
P_{\mu e} &=& B - 2xyfg + 4\hat A \epsilon_{e\tau} s_{23} c_{23} \{xf^2 + y g f\}\;.
\label{cond1}
\end{eqnarray}
If we take $\delta_{\rm CP}=270^\circ$ (which is close to $\delta_{\rm CP} = 240^\circ$) then $P(\nu_\mu \to \nu_e)$ will become,
\begin{eqnarray}
P_{\mu e} &=& B + 2xyfg + 4\hat A \epsilon_{e\tau} s_{23} c_{23} \{-xf^2 + y g f\}\;.
\label{cond2}
\end{eqnarray}
From the above two equations we obtain:
\begin{eqnarray}
P_{\mu e} (\delta_{\rm CP} = 90^\circ) - P_{\mu e} (\delta_{\rm CP} = 270^\circ) = 4 x f (2 \hat A \epsilon_{e \tau} s_{23} c_{23} f - yg)\;.
\end{eqnarray}
The above equation is true for neutrinos. For antineutrinos, similarly one can show that
\begin{eqnarray}
\overline{P}_{\mu e} (\delta_{\rm CP} = 90^\circ) - \overline{P}_{\mu e} (\delta_{\rm CP} = 270^\circ) = - 4 x f (2 \hat A \epsilon_{e \tau} s_{23} c_{23} f - yg)\;.
\end{eqnarray}
As $yg$ is always less than $2 \hat A \epsilon_{e \tau} s_{23} c_{23} f$,  probabilities for $\delta_{\rm CP}=90^\circ$ will be higher than probabilities for $\delta_{\rm CP}=270^\circ$ for neutrinos and opposite for antineutrinos. However, the value of $f$ is large for neutrinos and small for antineutrinos. This is why the red and the blue curves are largely separated in neutrinos and they are closely spaced in antineutrinos. This explains why the $\delta_{\rm CP}=90^\circ$ curve is free from hierarchy degeneracy for the case of neutrinos and degenerate with IH in antineutrinos. Note that the above conclusion is independent of the baseline.

\section{Summary and conclusions}
\label{sum}

In this paper we have studied the capability of P2O experiment to measure neutrino mass hierarchy and its sensitivity to NSI in neutrino propagation. P2O is a future proposed long-baseline experiment which will use a neutrino source located in Protvino, Russia and these neutrinos will be detected at the ORCA detector located in the Mediterranean sea. This experiment will have the longest possible baseline of 2595 km among all the other future long-baseline experiments. As the determination of mass hierarchy and NSI in neutrino propagation depends on the matter effect, in principle P2O should have highest sensitivity towards determination of neutrino mass hierarchy and constraining the NSI parameters. In addition, as the baseline of P2O is close to the bi-magic baseline, it has the capability to determine neutrino mass hierarchy without the hierarchy - $\delta_{\rm CP}$ degeneracy. In Ref. \cite{Akindinov:2019flp}, three possible designs of this experiment is studied: (i) minimal configuration of ORCA detector with 90 KW beam, (ii) updated accelerator configuration of 450 KW beam with ORCA detector and (iii) updated accelerator configuration of 450 KW beam with the updated Super-ORCA detector. In this work, we have considered the minimal configuration of P2O and compared its sensitivity with DUNE. First we showed that though the number of signal events for P2O are more as compared to DUNE, the hierarchy sensitivity of DUNE is better than P2O. This is because of the limited background rejection capability of P2O. The appearance channel signal events in P2O are reconstructed from the shower events where the backgrounds corresponding to $\nu_\mu$, NC and $\nu_\tau$ are very high as compared to the DUNE detector. One way to solve this problem is to reduce the backgrounds and improve the systematic errors in the appearance channel without increasing the beam power. We found that for the background reduction factor of 0.46 and background systematic error of 4\%, the sensitivity of P2O becomes comparable with DUNE for $\delta_{\rm CP} = 195^\circ$ which is the current best-fit value of this parameter. We call this configuration as optimized P2O. In general, the hierarchy sensitivity of the optimized P2O is equivalent (better) as compared to DUNE for $180^\circ (0^\circ) < \delta_{\rm CP} < 360^\circ (180^\circ)$ which is currently favoured (disfavoured) by the global analysis. This improved sensitivity as compared to ORCA can be achieved in the future Super-ORCA configuration which will be a 10 times denser detector geometry compared to ORCA. In Super-ORCA, the efficiency for selecting the signal events and rejecting the background events will be higher. For the study of NSI we considered the NSI parameters $\epsilon_{e \mu}$ and $\epsilon_{e \tau}$. Our choice of NSI parameters are motivated by recent study of Refs. \cite{Chatterjee:2020kkm,Denton:2020uda} where it was shown that the discrepancy between the $\delta_{\rm CP}$ measurement in the experiments T2K and NO$\nu$A can be resolved by introducing the NSI parameters $\epsilon_{e \mu}$ and $\epsilon_{e \tau}$. For our study we have taken the best-fit values of Ref. \cite{Chatterjee:2020kkm} as a reference. Our analysis has shown that for $\epsilon_{e\mu}$ the sensitivity of DUNE is better than optimized P2O if we do not include $\epsilon_{e\tau}$ in the analysis whereas they become similar if we include $\epsilon_{e\tau}$ in the analysis. However, for $\epsilon_{e\tau}$ the sensitivity of DUNE is always better than optimized P2O irrespective of inclusion of $\epsilon_{e\mu}$ in the analysis. From our analysis we noticed that the values $|\epsilon_{e\mu}| = 0.15$ and $|\epsilon_{e\tau}| = 0.27$ (the best-fit values of the NSI parameters as obtained in Ref. \cite{Chatterjee:2020kkm}) will be excluded at 90\% C.L. by all the experiments irrespective of the values of $\phi_{e \mu}$ and $\phi_{e \tau}$ except P2O when both $\epsilon_{e\mu}$ and $\epsilon_{e\tau}$ are included in the analysis. For study of hierarchy sensitivity in the presence of NSI we used the true value of NSI parameters as obtained in Ref. \cite{Chatterjee:2020kkm}. The hierarchy sensitivity in presence of NSI is lower than sensitivity in the standard three flavour scenario for $\delta_{\rm CP} = 270^\circ$ and higher than the sensitivity in the standard three flavour scenario for $\delta_{\rm CP} = 90^\circ$. The change of hierarchy sensitivity due to NSI is higher in P2O as compared to DUNE with respect to the hierarchy sensitivity in the standard three flavour scenario. As in the standard case, in presence of NSI, the hierarchy sensitivity of optimized P2O  is not better than DUNE for the current favourable values of $\delta_{\rm CP}$ which is $180^\circ < \delta_{\rm CP} < 360^\circ$ as obtained by the global analysis.
 We have also identified a degeneracy between NH and ($\delta_{CP} \sim 270^\circ$, $\phi_{e\tau} \sim 270^\circ$) with IH in presence of $\epsilon_{e \tau}$ in the neutrino probabilities. For antineutrino probabilities, both $\delta_{CP} \sim 270^\circ$ and $\delta_{CP} \sim 90^\circ$ in NH are degenerate with IH. This degeneracy is independent of the baseline length. Because of this the hierarchy sensitivity became almost zero for $\delta_{CP} \sim 270^\circ$ and non-zero for $\delta_{CP} \sim 90^\circ$. However, if we had considered only antineutrino run, then hierarchy sensitivity would have vanished for both $\delta_{CP} \sim 270^\circ$ and $\delta_{CP} \sim 90^\circ$. In addition we have shown that the hierarchy sensitivity of optimized P2O in presence of the NSI parameter $\epsilon_{e \mu}$ is equivalent to the sensitivity of the upgraded P2O with a 450 KW beam.

In conclusion, we want to state that in this present work we considered the minimal configuration of P2O and proposed an optimized configuration of P2O in terms of background and systematic errors without improving the beam power. This configuration can be achieved in the Super-ORCA configuration in future. The sensitivity of the optimized P2O is similar to the sensitivity of P2O with the upgraded beam. However, the best sensitivity of P2O can be achieved with a upgraded beam and Super-ORCA configuration. The results presented in this work are important to understand the effect of background and systematic errors in P2O for the determination of mass hierarchy and also to show its sensitivity in presence of NSI.

\section*{Acknowledgements}

DKS acknowledges Prime Ministers Research Fellowship, Govt. of India. MG acknowledges Ramanujan Fellowship of SERB, Govt. of India, through grant no: RJF/2020/000082. RM acknowledges the support from University of Hyderabad through the IoE project grant IoE/RC1/RC1-20-012 and  SERB, Government of India, through grant No. EMR/2017/001448. We acknowledge the use of CMSD HPC facility
of Univ. of Hyderabad to carry out computations in this work. We also thank Papia Panda and Priya Mishra for useful discussions. 

\section*{Appendix}

\begin{figure}[hbt]
\begin{center}
\includegraphics[width=0.5\textwidth]{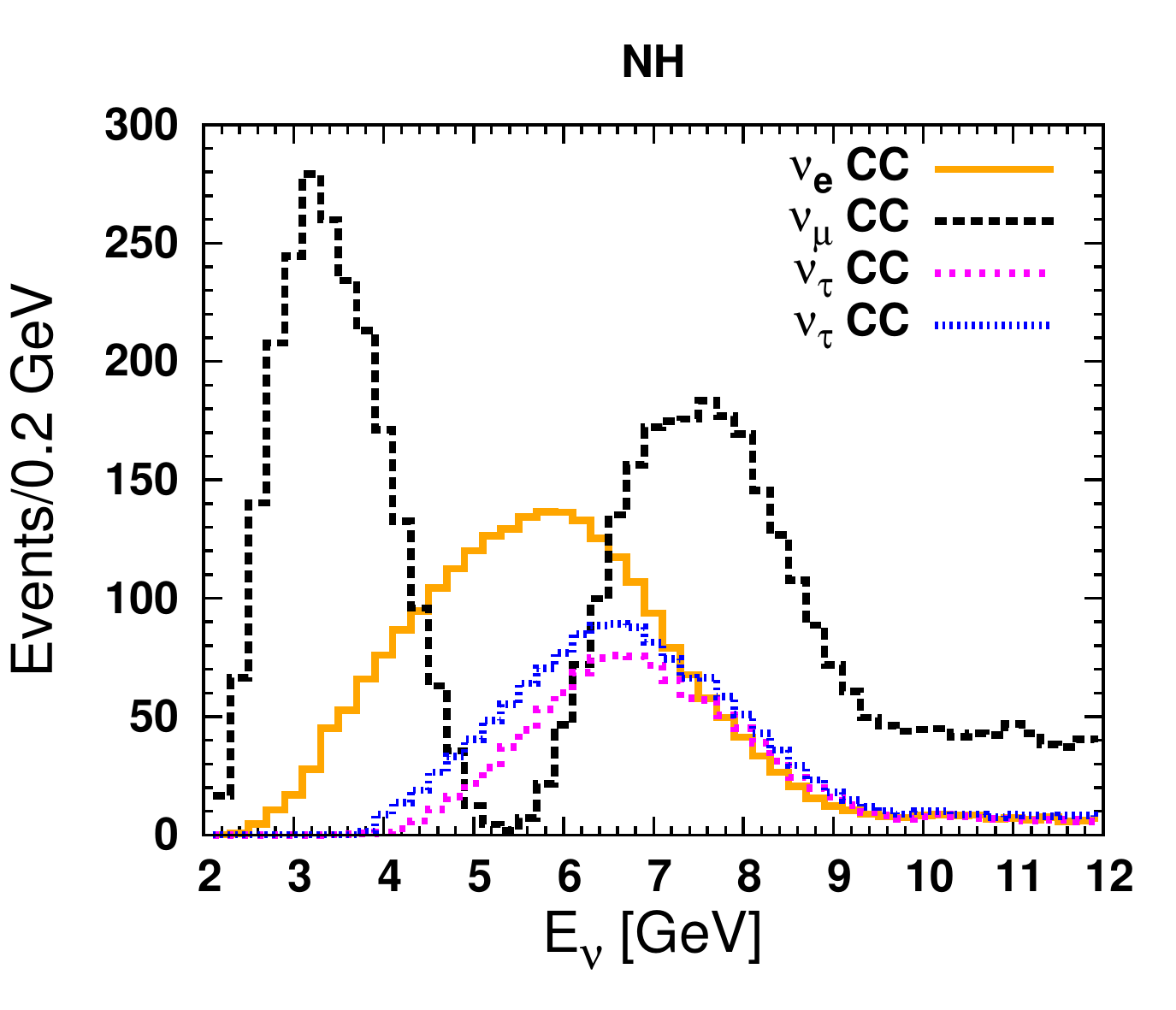}
\includegraphics[width=0.49\textwidth]{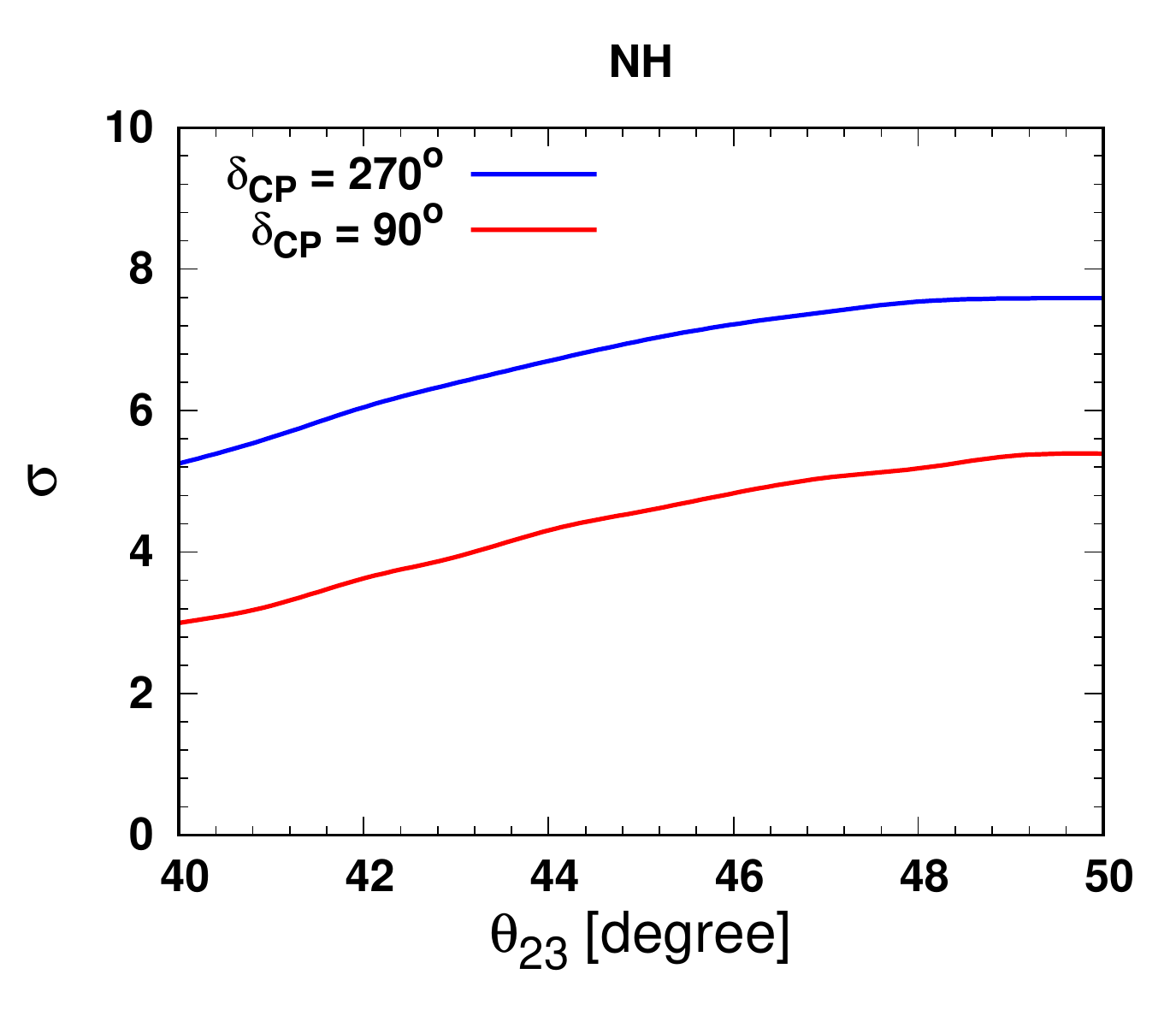}
\end{center}
\caption{event spectrum of P2O as a function true energy (left panel) and the hierarchy sensitivity of P2O as a function of $\theta_{23}$ (right panel).}
\label{fig_appendix}
\end{figure}

In this appendix, we give the event spectrum of P2O as a function true energy and the hierarchy sensitivity of P2O as a function of $\theta_{23}$. We have generated Fig. \ref{fig_appendix} for normal hierarchy of the neutrino masses. In generating this figure, we have used the specifications and the value of oscillation parameters as given in \cite{Akindinov:2019flp}. This figure shows that our sensitivity matches with the sensitivity of P2O as given in  \cite{Akindinov:2019flp}.

\bibliographystyle{JHEP}
\bibliography{nsi_P2O}
  
\end{document}